\documentclass[11pt,hyper,a4paper]{JHEP}
\usepackage{amsmath,graphicx,epsfig,amssymb,multirow}
\allowdisplaybreaks
\newcommand{\ba}{\begin{array}}
\newcommand{\ea}{\end{array}}
\newcommand{\beq}{\begin{equation}}
\newcommand{\eeq}{\end{equation}}
\newcommand{\comment}[1]{}
\def\bt{\begin{table}}
\def\et{\end{table}}
\def\bc{\begin{center}}
\def\ec{\end{center}}
\def\bi{\begin{itemize}}
\def\ei{\end{itemize}}
\def\bea{\begin{eqnarray}}
\def\eea{\end{eqnarray}}
\def\beas{\begin{eqnarray*}}
\def\eeas{\end{eqnarray*}}

\def\N0{\widetilde{\chi}^0}

\catcode`@=11 
\def \gsim{\mathrel{\mathpalette\@versim>}}
\def \lsim{\mathrel{\mathpalette\@versim<}}
\def \@versim#1#2{\lower0.4ex\vbox{\baselineskip\z@skip\lineskip\z@skip
     \lineskiplimit\z@\ialign{$\m@th#1\hfil##\hfil$%
     \crcr#2\crcr\sim\crcr}}}
\catcode`@=12 

\preprint{HIP-2010-24/TH}
\title{\boldmath On measurement of  top polarization as a probe of  $t \bar t$
production mechanisms at the LHC}
\author{Rohini M. Godbole\thanks{Permanent address:  Center for 
High Energy Physics, Indian Institute of Science, Bangalore 560 012,
India.}\\Theory Unit, CERN, CH-1211, Geneva 23,
Switzerland\\
Email: \email{rohini@cts.iisc.ernet.in}}
\author{Kumar Rao\\
Department of Physics, University of Helsinki and Helsinki Institute of
Physics, \\ P.O. Box 64, FIN-00014, Helsinki,
Finland\\Email: \email{kumar.rao@helsinki.fi}}
\author{Saurabh D. Rindani\\
Theoretical Physics Division, Physical Research Laboratory,
Navrangpura, Ahmedabad~380 009, India\\
Email: \email{saurabh@prl.res.in}}
\author{
Ritesh K. Singh\\
Institut f\"ur Theoretische Physik und Astronomie, Universit\"at 
W\"urzburg, 97074, Germany\\
Email: \email{singh@physik.uni-wuerzburg.de}}
\abstract{

\def\lessthansquiggle{\raise.3ex\hbox{$<$\kern-.75em\lower1ex\hbox{$\sim$}}}

In this note we demonstrate the use of top polarization in the study of
$t \bar t$ resonances at the LHC, in the possible case where the dynamics
implies a non-zero top polarization. 
As a probe of top polarization we construct 
an asymmetry in the decay-lepton azimuthal angle distribution 
(corresponding to the sign of  $\cos\phi_\ell$) in the laboratory. 
The asymmetry is non-vanishing even for a symmetric collider like the
LHC, where a positive $z$ axis is not uniquely defined.  The angular 
distribution of the leptons has the advantage of being a faithful top-spin 
analyzer, 
unaffected by possible anomalous $tbW$ couplings, to linear order.  We study, 
for purposes of demonstration, the case of a $Z'$ as might exist in the little 
Higgs models. We identify kinematic cuts which ensure that our asymmetry 
reflects the polarization in sign and magnitude.
We investigate possibilities at the LHC
with two energy options: $\sqrt{s} = 14$ TeV and 
$\sqrt{s} = 7$ TeV,  as well as at the Tevatron.
At the LHC the model predicts net top quark polarization 
of the order of a few per cent for $M_{Z'} \simeq 1200 $ GeV, being as high as 
$10 \%$ for a smaller mass of the $Z'$ of $700$ GeV and for the largest allowed 
coupling in the model, the values being higher for the $7$ TeV option. These 
polarizations translate to a deviation from the standard-model value of 
azimuthal asymmetry of up to about 
$4\%$ ($7\%$) for $14$ ($7$) TeV LHC, whereas for the Tevatron, values as high 
as $12\%$ are attained. For the $14$ TeV LHC with an integrated
luminosity of 10 fb$^{-1}$, these
numbers translate into a $3 \sigma$ sensitivity over a large part of the
range $500 \lesssim M_{Z'} \lesssim 1500 $ GeV. 

}
\keywords{Top polarization, top pair production, LHC, Tevatron }

\begin{document}

\section{Introduction} \label{sec-1}
The properties and interactions of quarks and leptons belonging to
the third family are still relatively poorly known. Universality of 
interactions of all the three generations is a natural prediction of
the Standard model (SM), but the number of generations and the
relative masses in the model seem completely ad hoc. Serious
constraints have been set on the universality of couplings of
the first two generations, but for the third, it is less well tested.
The closeness of the top quark mass to the Electroweak symmetry breaking (EWSB)
scale, in fact, leads to speculations that  it might be closely related
to an answer to the as yet unsolved problem of the EWSB and alternatives to the SM Higgs mechanism almost always involve the top
quark ~\cite{Hill:2002ap}. Most of the Beyond the Standard Model (BSM) 
scenarios have a new particle which is closely related to the top quark 
in one way or the other and hence the top quark always plays an important role
in BSM searches at colliders:  be it the supersymmetric 
partner of the top (the stop)~\cite{susybook} or the heavy top expected in 
the Little Higgs models~\cite{ArkaniHamed:2001ed,ArkaniHamed:2001nc}.  
In addition, many BSM models also predict strongly coupled $t \bar t$ resonances,
with or without preferential  couplings to a $t \bar t$ pair
\cite{ArkaniHamed:2001ed,ArkaniHamed:2001nc,Frampton:1987dn,Frampton:1987ut,Chivukula:1996yr,Randall:1999ee,Kilic:2008pm}. Clearly, one expects a top factory
such as the LHC to be an ideal place to hunt for BSM physics in top 
production~\cite{Beneke:2000hk,Bernreuther:2008ju,Han:2008xb,Frederix:2007gi}.
Already at the Tevatron, the study of top physics has proved quite fruitful with
combined fits providing constraints on  masses and production cross-sections of 
$t \bar t$ resonances~\cite{:2007dz,:2007dia,Abazov:2008ny,Aaltonen:2009tx} 
as well as from consideration of their contribution to the total $t \bar t$ 
cross-section~\cite{Guchait:2007ux,Choudhury:2007ux}. The observation of a
forward-backward asymmetry~\cite{:2007qb,Aaltonen:2008hc} in $t \bar t$ 
production, differing from the SM expectation at  more than $2 \sigma$ level,
is perhaps one of the few `disagreements' between the experimental data and the
SM predictions and has found a host of BSM explanations.

With a large mass of  about 175 GeV, the top quark has an extremely short 
lifetime, calculated in the SM to be $\tau_t=1/\Gamma_t \sim 5 \times 10^{-25}$
second. This is an order of magnitude smaller than the hadronization time 
scale, which is roughly $1/\Lambda_{\rm{QCD}} \sim 3 \times 10^{-24}$ second.
Thus the top decays before it can form bound states with lighter quarks. As a 
result the kinematical distribution of its  decay products retain the memory 
of the top spin direction. Clearly, top spin information holds more clues to 
top production dynamics than just the cross-section. For example, in the MSSM,
the expected polarization of the top produced in the decay of the stop 
can  provide information on model parameters such as mixing in the
neutralino/sfermion sector or amount of 
CP violation~\cite{Boos:2003vf,Gajdosik:2004ed}. Use of top polarization as 
a probe of  additional contributions to $t \bar t$ production due to sfermion 
exchange in R-parity violating MSSM, was suggested in 
Ref.~\cite{Hikasa:1999wy}. 
It is interesting to note that this would also imply a 
forward-backward asymmetry in top production such as reported 
at the Tevatron. Thus in this case top polarization may be able to provide 
a discrimination between different explanations that have been put forward.
More generally, top polarization can offer
separation between different processes responsible for top production 
\cite{Arai:2010ci} or can allow discrimination between 
different BSM models with differing spins of the 
top partner~\cite{Perelstein:2008zt,Shelton:2008nq}.  

Probing BSM dynamics in top physics can thus receive an additional boost if top
polarization or $t \bar t$ spin-spin correlations can be faithfully inferred 
from the kinematic distributions of its decay products. For example, expected
kinematic distributions of the decay products of the top have been used to
fine tune search strategies for BSM physics such as the top partner in Little 
Higgs models or the Kaluza-Klein (KK) gluons expected in brane-world models
with warped extra dimensions~\cite{Nojiri:2008ir,Agashe:2006hk,Djouadi:2007eg}.
The large Yukawa coupling of the $t$ quark with the Higgs boson makes it an 
ideal candidate for studying properties of the Higgs boson, particularly so 
because it can offer a way to distinguish between the chirality conserving 
gauge interactions and chirality flipping Yukawa interactions. In fact, the 
final state top quark polarization for associated $t \bar t H$ production in 
$e^+e^-$ collisions can reflect the CP-parity of the Higgs 
boson~\cite{BhupalDev:2007is}. For hadronic $t\bar{t}$ production, 
spin-spin correlations between the decay leptons from the $t$ and $\bar{t}$ 
have been extensively studied in the SM and for BSM
scenarios \cite{Beneke:2000hk,spincorr}. These
spin-spin correlations measure the asymmetry between the production of like and
unlike helicity pairs of $t\bar{t}$ which can probe new physics in top pair
production. Correlations between the spins of $t , \bar t$ produced in
the decay of 
the Higgs or in association with the Higgs, also reflect the spin-parity of 
the Higgs boson~\cite{Bernreuther:2008ju}. Strategies have been outlined for 
using these correlations for studying KK graviton excitations 
\cite{Arai:2007ts} as well. However, measuring spin correlations requires the 
reconstruction of the $t$ and $\bar{t}$ rest frames, which is difficult, if 
not impossible, at the LHC. 
In this note we wish to explore use of {\it single top polarization} as a 
qualitative and quantitative probe of new physics in $t  \bar t$ production,
keeping in mind that it would offer higher statistics compared to studies
of spin-spin correlations.

As has been noted already, most of the spin studies mentioned here and various
suggestions for similar studies always involve the 
construction of observables in the rest frame of the decaying top quark. It 
would be interesting and useful to construct observables to track the decaying 
top quark polarization using kinematic variables in the laboratory frame. 
It is well known that the angular distribution of a decay fermion, measured 
with respect to the direction of the top spin in the rest frame of the top 
quark, can be used as a top spin analyzer, the lepton being the most efficient 
analyzer. 
This angular distribution translates into specific kinematic distributions
for the decay lepton in the laboratory frame where the $t$ is in motion, 
depending on the polarization of the decaying quark. 
However, it is the energy averaged angular distribution of the decay lepton
which is found to be independent of any possible anomalous contribution to the 
$tbW$  vertex~\cite{hikasa,rindani,ritesh1,ritesh2}. If the dynamics gives rise
to net polarization of the decaying top quark, at a collider like the Tevatron 
this can translate into a polarization asymmetry with respect to the beam 
direction and hence an asymmetry in the decay lepton angular distribution 
with respect to (say) the proton direction in the laboratory. 
However, at a collider like the LHC, where the direction of either proton can
be chosen to be the positive direction of the $z$ axis, simple observables 
like this will vanish even if the dynamics gives rise to a polarization
asymmetry for  the top (and hence an angular asymmetry of the decay lepton
in the laboratory) with respect to the direction of one of the protons. Hence, 
it is necessary to construct a non-vanishing observable which will faithfully 
reflect such non-zero polarization.

In this note we address the issue of constructing such an observable at the LHC
which would serve as a faithful measure of the polarization of the top quark
arising from  the dynamics of the subprocess of production. We show that
it is possible to construct an asymmetry, measured in the laboratory frame, 
using the distribution in the azimuthal angle of the decay lepton with respect 
to the $x-z$ plane, being the plane containing the direction of one of the 
protons as the $z$ axis and the direction of the decaying $t$ quark. This 
observable 
directly reflects the sign and the magnitude of the $t$ polarization, with a 
suitable choice of kinematic cuts.  We demonstrate this using as an example 
the  production  of a $t \bar t$ resonance, with chiral couplings to the 
fermions, as in the Littlest Higgs 
Model~\cite{ArkaniHamed:2001ed,ArkaniHamed:2001nc}. A preliminary
study of the possibility  of using this observable and hence the top 
polarization to get information on the structure of couplings of these 
resonances with the $t/\bar t$, has been presented 
elsewhere~\cite{usleshouch,Djouadi:2007eg,Godbole:2009dp}.

In Section 2 we present the details of the model as well as the 
calculational framework. In Section 3 we present results.
We begin by showing our results for $t$ polarization at the 
LHC, both for $\sqrt{s} = 14$ and $7$ TeV,  for $Z'$ production with 
chiral couplings expected in the Littlest Higgs Model, over the 
parameter space of the model, with and without integration over the invariant
mass of the $t \bar t$ pair. We then describe the construction
of the azimuthal angle asymmetry in the laboratory frame 
as a measure of the $t$ polarization. Next  we show its  dependence on the
kinematic variables in the problem. This then help us identify  the kinematic
cuts, such that this asymmetry  reflects the size and the sign of the $t$
polarization faithfully. We then present our results on the sensitivity of the 
LHC at $14$ and $7 $ TeV, as well as that for the Tevatron, for the $Z'$ model 
under consideration and then conclude.

\section{Model and calculational framework}
There exist various examples of $t \bar t$ resonances in different BSM 
scenarios; the strongly interacting ones, like KK gluons, colorons, 
axigluons,  as well as  various other versions of additional $Z'$ 
resonances~\cite{Langacker:2008yv} that occur in almost all the BSM scenarios, 
with a variety of couplings to different fermions. In the former case of 
strongly interacting resonances, there are two classes, one with enhanced 
couplings to $t \bar t$ pair which includes KK gluons and the other without 
such enhanced couplings, which includes colorons, axigluons etc.   
Search for additional $Z'$ in the leptonic channel
is an item with high priority on the agenda of the LHC  first 
run~\cite{Langacker:2009su,Salvioni:2009jp,Salvioni:2009mt}.
In the leptonic channel, even with the $1$ fb$^{-1}$ luminosity and lower
centre of mass energy of $7$ TeV, the LHC in this first run should be able
to probe beyond the current Tevatron limits~\cite{Amsler:2008zzb}. 
Since couplings to the third generation of fermions could be substantially 
different in different models, even if we are blessed with an early discovery 
of a $Z'$ in the leptonic channel at the LHC, a clear and complete 
characterization of such a resonance and hence the BSM physics it may 
correspond to, will require determination of these. 
A  $Z'$ with a chiral coupling to $t \bar t$ would give rise to 
substantial polarization of the top, which could be a distinguishing feature of 
the model. Here we illustrate how azimuthal distributions can be used to 
investigate top polarization, using the example of a $Z'$ with purely 
chiral couplings, such as the one that occurs in a model similar to the 
Littlest Higgs model. 

We consider a $Z'$ of mass $M_{Z'}$ whose couplings to quarks are purely 
chiral, given by~\cite{hanLH}
\begin{equation}\label{zprimeff}
\mathcal L_{q\bar q Z'} = -\frac{1}{2}g \cot(\theta) \sum_{i=1} \left[ \bar u_i \gamma_\mu P_{L,R} u_i
	- \bar d_i \gamma_\mu P_{L,R} d_i \right] Z'^\mu,
\end{equation}
where $g$ is the weak coupling constant and $\cot(\theta)$ is a free parameter 
in the model. The subscripts $P_{L,R}$ refer respectively to left- and 
right-chiral projection operators.  If $Z'$ is $Z_H$ of the Littlest Higgs 
model, we would choose the subscript $L$ in the above equation. However, we 
will also use for illustration a model in which $Z'$ has pure right-chiral 
couplings, for which we choose the subscript $R$. With the couplings of 
eq. (\ref{zprimeff}), the total decay width of $Z'$ comes out to be
\begin{equation}\label{zpwidth}
\Gamma_{Z'} = \frac{g^2}{96\pi}M_{Z'}\cot^2(\theta)\left[21 +
 3\sqrt{1-4m_t^2/M_{Z'}^2}(1-m_t^2/M_{Z'}^2)
\right],
\end{equation}
where the partial decay widths into $W^{+}W^{-}$ and $ZH$ have been
neglected~\cite{hanLH}.
Since the dominant production mechanism for $t \bar t$ in the SM is
parity-conserving,
top polarization expected in the SM is very small, both at the Tevatron and the
LHC.
However, a $Z'$ with chiral couplings as given by eq. (\ref{zprimeff}) can give
rise to substantial top and anti-top polarization, 
for sufficiently large values of 
$\cot (\theta)$ and for values of $m_{t \bar t}$ comparable to $M_{Z'}$.
The kinematic distribution of the decay fermions coming from the 
$t$ or $\bar t$  
can be used to get information on this polarization. Below we first discuss 
how this is accomplished in the rest frame of the decaying top and also
sketch out the necessary formalism used to calculate the correlated production 
and decay of the top keeping the spin information.

The  angular distribution of the  decay products of the top is correlated with
the direction of the top spin.  In the SM, the dominant decay mode is 
$t\to b W^+$, with the $W^+$ subsequently decaying to $l^+ \nu_{\ell}$, $u \bar{d}$
or $c\bar{s}$, with $l$ denoting any of the leptons. 
For a top quark ensemble  with polarization $P_t$ , in the top rest frame 
the angular distribution of the decay product $f$ (denoting $W^+$, $b$,
$\ell^+$,
$\nu_{\ell}$, $u$ and $\bar d$ ) is given by,
\begin{equation}
\frac{1}{\Gamma_f}\frac{\textrm{d}\Gamma_f}{\textrm{d} \cos \theta _f}=\frac{1}{2}(1+\kappa _f P_t \cos \theta _f).
\label{topdecaywidth}
\end{equation}
Here $\theta_f$ is the angle between the decay product  $f$ and the top spin 
vector in the top rest frame, and the degree of top polarization $P_t$ for the 
ensemble is given by
\begin{equation}
 P_t=\frac{N_\uparrow - N_\downarrow}{N_\uparrow + N_\downarrow}
\label{ptdef}
\end{equation}
where $N_\uparrow$ and $N_\downarrow$ refer to the number of positive and 
negative helicity tops respectively. $\Gamma_f$  denotes  the partial 
decay width.  $\kappa_f$ is a constant which depends on the weak isospin
and the mass of the decay product $f$ and is called its spin
analyzing power.  Obviously, a larger value of $\kappa_f$ makes  $f$ a more
sensitive probe of the top spin. At tree level, the charged lepton and $d$
anti-quark are thus best spin analyzers with $\kappa_{l^+}=\kappa_{\bar{d}}=1$,
while $\kappa_{\nu_{\ell}}=\kappa_{u}=-0.31$, with 
$\kappa_{b} = -\kappa_{W^+}=-0.41$. eq.~(\ref{topdecaywidth}) thus tells us 
that 
the $l^+$ or $d$ have the largest probability of being emitted in the direction
of the top spin and the least probability in the direction opposite to the 
spin. Since among these two it is the charged lepton ($l^+$) for which the 
momenta can
be determined with  high precision, one usually focuses on the semi-leptonic
decay of the $t$ (corresponding to the leptonic decay of the $W^+$), for spin
analysis. 

Since the values of $\kappa_f$ in 
eq.~(\ref{topdecaywidth}) follow from the $V$-$A$ structure of the 
$W f \bar f'$ 
couplings, it is important to consider how they are affected by a nonzero
anomalous $tbW^+$ coupling.  New physics may appear in the $tbW$ decay vertex,
apart from that in top production, leading to changed decay width and 
distributions for the $W^+$ and $l^+$. A model-independent form for 
the $tbW$ vertex can be written as
\begin{equation}
\Gamma^\mu =\frac{-ig}{\sqrt{2}}\left[\gamma^{\mu}(f_{1L} P_{L}+f_{1R}P_{R})
-\frac{i \sigma^{\mu \nu}}{m_W}(p_t -p_b)_{\nu}(f_{2L}P_{L}+f_{2R}P_{R})\right]
\label{anomaloustbW}
\end{equation}
where for the SM $f_{1L}=1$ and the anomalous couplings 
$f_{1R}=f_{2L}=f_{2R}=0$.
Luckily, as has been shown in Ref.~\cite{ritesh2}  and will be discussed 
briefly below, it is precisely for the two best spin analyzers, the $l^+$ 
and the $\bar d$, that the value of $\kappa_f$ in
eq.~(\ref{topdecaywidth}) 
remains unchanged to leading 
order in the anomalous couplings. Hence this distribution is indeed a 
robust top spin analyzer.

So far we have discussed the somewhat academic issue of the correlation 
between the direction of the top spin, in an ensemble with degree of 
polarization $P_t$, and the angular distribution of the charged decay 
lepton. However, in an actual
experiment we have to consider the process of top production and its 
semi-leptonic decay and perform the calculation  preserving information on 
the top spin from production to decay. To this end, let us consider a 
generic process of top pair production and subsequent semi-leptonic decay 
of $t$ and inclusive decay of $\bar{t}$, 
$A B \to t \bar{t} \to b \ell^+ \nu_\ell X$. Since 
$\Gamma_t/m_t \sim 0.008$, we 
can use the narrow width approximation (NWA) to write the cross section as 
a product of the $2\to 2$ production cross section times the decay 
width of the top. To preserve coherence of the top 
spin in production and decay, we need to use the spin 
density matrix formalism. The amplitude squared can be
factored into production and decay parts in the NWA~\cite{ritesh2}:
\begin{eqnarray}
\overline{|{\cal M}|^2} = \frac{\pi \delta(p_t^2-m_t^2)}{\Gamma_t m_t}
\sum_{\lambda,\lambda'} \rho(\lambda,\lambda')\Gamma(\lambda,\lambda').
\label{matelsq}
\end{eqnarray}
where $\rho(\lambda,\lambda')$ and $\Gamma(\lambda,\lambda')$ are the 
$2 \times 2$ top production and decay spin density matrices respectively, with 
$\lambda,\lambda' =\pm 1$ denoting the top helicity. The phase space 
integrated $\rho(\lambda,\lambda')$ gives the  polarization density matrix 
and can be parametrized as
\begin{eqnarray}
\sigma(\lambda,\lambda^\prime)=
\frac{\sigma_{\rm{tot}}}{2}\left(
\begin{tabular}{cc}
$1+\eta_3$ & $\eta_1 - i\eta_2$ \\ 
$\eta_1 + i\eta_2$ & $1-\eta_3$
\end{tabular} \right),
\label{poldm}
\end{eqnarray}
The (1,1) and (2,2) diagonal elements are the cross sections 
for the production of positive and negative helicity tops.   
$\sigma_{\rm{tot}}=\sigma(+,+)+\sigma(-,-)$ gives the total cross section, 
whereas the difference $\sigma_{\rm{pol}} = \sigma(+,+) - \sigma(-,-)$ 
is the polarization dependent part of the cross-section. In fact 
$\eta_{3}$ is the degree of {\it longitudinal} polarization and is 
given by the ratio of  $\sigma_{\rm pol}$ to $\sigma_{\rm tot}$ as,
\begin{equation}
\eta_3 = P_t = \frac{\sigma(+,+)-\sigma(-,-)}{\sigma(+,+)+\sigma(-,-)} =
\frac{\sigma_{\rm pol}}{\sigma_{\rm{tot}}}.
\label{eta3def}
\end{equation}
The production rates of the top with {\it transverse} polarization are 
given by the  off-diagonal elements involving $\eta_1$ and $\eta_2$ , the two
being the transverse components of the top polarization parallel and 
perpendicular to the production plane respectively. These are given by, 
\begin{eqnarray}
 \eta_1=\frac{\sigma(+,-)+\sigma(-,+)}{ \sigma(+,+)+\sigma(-,-)},\quad
 i\eta_2=\frac{\sigma(+,-)-\sigma(-,+)}{ \sigma(+,+)+\sigma(-,-)}.
\end{eqnarray}

The spin dependence of the top decay is included via the top decay density 
matrix of eq.~(\ref{matelsq}), $\Gamma(\lambda,\lambda')$.  For the process 
$t\to b W^{+}\to b \ell^{+} \nu_\ell$ this can be written in a 
Lorentz invariant form as
\begin{equation}
\Gamma(\pm,\pm)=\frac{2 g^4}{|p_W ^{2} -m_{W}^{2}+i \Gamma _{W} m_{W}|^{2}}
(p_b \cdot p_\nu)
 \left[(p_\ell \cdot p_t) \mp m_t (p_\ell \cdot n_3)\right],
 \label{eq:ddd}
\end{equation}
for the diagonal elements and
\begin{equation}
\Gamma(\mp,\pm)=-\frac{2 g^4}{|p_W ^{2} -m_{W}^{2}+i \Gamma _{W} m_{W}|^{2}} 
\,m_t \,\,(p_b \cdot p_\nu) \,\, p_\ell \cdot (n_1 \mp i n_2),
\label{eq:ddo}
\end{equation}
for the off-diagonal ones. Here the $n^{\mu}_{i}$'s ($i=1,2,3$) are the 
spin 4-vectors for the top with 4-momentum $p_t$, with the properties 
$n_i \cdot n_j =-\delta_{ij}$ and $n_i \cdot p_t =0$. 
For decay in the rest frame they take the standard form $n^{\mu}_{i}=(0, 
\delta_{i}^{k})$. As shown in \cite{ritesh2}, in the rest frame of the $t$ 
quark the expression for $\Gamma(\lambda,\lambda')$, 
after phase space integration over the $b$ quark and $\nu_{\ell}$  momenta, 
factorizes into a lepton energy dependent part $F(E_\ell^0)$ and a
a function  $A(\lambda,\lambda')$ which depends only on the angles of
the decay lepton $\ell$:
\begin{eqnarray}
\langle\Gamma(\lambda,\lambda')\rangle= (m_t E_\ell^0) \ |\Delta(p_W^2)|^2 \
g^4 A(\lambda,\lambda') \  F(E_\ell^0).
\label{topdecaymatrix}
\end{eqnarray}
Here angular brackets denote an average over the azimuthal angle
of the $b$ quark w.r.t the  plane of the $t$ and the $\ell$ momenta and
$\Delta(p_W^2)$ stands for the propagator of the $W$.
The azimuthal correlation between $b$ and $\ell$ is 
sensitive to new physics in the $tbW$ couplings; this averaging eliminates 
any such dependence.  
Using the NWA for the top and the result of eq.~(\ref{topdecaymatrix})
the differential cross section for top production and decay, after 
integrating over $\nu_{\ell}$ and $b$, can be written as
\begin{eqnarray}
d\sigma&=&\frac{1}{32 \ \Gamma_t m_t} \ \frac{1}
{(2\pi)^4} \left[ \sum_{\lambda,\lambda'} d\sigma (\lambda,\lambda') \
\times \ g^4 A(\lambda,\lambda') \right] \  |\Delta(p_W^2)|^2\\ \nonumber
&\times& d\cos\theta_t \ d\cos\theta_\ell \ d\phi_\ell ~
E_\ell \ F(E_\ell) \ dE_\ell \ \ dp_W^2.
\label{dsigell}
\end{eqnarray}

As shown in \cite{ritesh2}, all three  components of the top
polarization,  $\eta_i,i =1,3$, can be  extracted by 
a suitable combination of lepton polar and azimuthal angular asymmetries,
constructed  by measuring the angular distributions 
of the decay lepton in the top rest frame. For example, as is expected  from 
eq.~(\ref{topdecaywidth}), $\eta_3 = P_t$ is simply given by a forward
backward asymmetry in the polar angle of the decay $\ell$ in the rest frame
of the top, with the $z$ axis along the top spin direction.  
Of course this requires reconstructing the top rest frame.
As pointed out in the introduction, it would be interesting to 
devise variables for the decay lepton in the lab frame, which can be easily 
measured and are sensitive to top polarization.

The factorization of the $\langle\Gamma(\lambda,\lambda')\rangle$
of  eq.~(\ref{topdecaymatrix}) into  $A(\lambda,\lambda')$ ,
a function only of the polar and azimuthal angles of $\ell$ in the
rest frame and $F(E_\ell^0)$ which is a function only of its energy $E_\ell^0$,
is very significant.  This factorization in fact leads to the result, 
mentioned already, viz. the energy averaged and normalized  decay lepton 
angular distribution (and also for the $\bar d$ quark), is independent of the 
anomalous $tbW$ coupling to the linear order. This has been shown very
generally for a $2\to n$ process, using NWA for the top  and neglecting
terms quadratic in the anomalous couplings in (\ref{anomaloustbW}) assuming 
new physics couplings to be small (for details see \cite{ritesh2}). 
This thus implies that the charged lepton angular distribution 
eq. (\ref{dsigell}) is a very robust probe of top polarization, 
free from any possible modification of the $tbW$ vertex due to new 
physics effect. Thus a measurement of top polarization via the angular
observables of the decay lepton, gives us a pure probe of new physics in 
top production process alone.  In
contrast, the energy distributions of the $l^+$ or the angular distributions of
the $b$ and $W$ may be  ``contaminated'' by the anomalous $tbW$ vertex, should
the new physics being probed contribute to that as well.

In the next section we study the azimuthal distribution of the decay 
charged lepton from a top quark in $t\bar t$ pair production at the  LHC 
in a model with $Z'$ with chiral couplings. We then define an azimuthal 
asymmetry sensitive to top polarization using this distribution.

\section{Results} 
We aim to investigate first the features of top polarization in the
presence of a $Z'$ resonance with chiral couplings. We will then 
examine the azimuthal distribution of charged leptons from top decay,
and a certain azimuthal asymmetry to be defined later, as a probe of top
polarization in the context of our chosen model.

For our numerical calculations we use {\tt CTEQ6L1} parton distributions 
with a scale $Q=m_t=175$ GeV.  To account for non-leading order contributions, 
we assume the $K$-factor for the entire process to be the same as that for 
the SM $t \bar t$ production and thus use a value of $1.40$ for LHC operating 
at $\sqrt{s} = 14$ TeV and $7$ TeV~\footnote{
We have checked using the tool HATHOR~\cite{Aliev:2010zk} that for {\tt CTEQ6M}
distributions the $K$-factor at $\sqrt{s} = 7$ TeV is the same as that for 
$\sqrt{s} = 14$ TeV. Hence  we use the value of $1.40$ for $\sqrt{s} = 7$ TeV 
in our case as well.  It should be noted however, that since we construct 
asymmetries, those results will not depend on the assumed $K$-factor, 
except the ones on the sensitivity reach that is possible using these 
asymmetries.} and $1.08$ for the Tevatron~\cite{Campbell:2006wx}.

\subsection{Top polarization}
To get an idea of the longitudinal top polarization that the production
of $Z'$ may give rise to, we begin by calculating the distributions of 
$\sigma_{\rm tot}$ and $\sigma_{\rm pol}$ in the $t \bar t$ invariant mass 
$m_{t\bar t}$.  Fig.~\ref{fig:mttLHC}  shows these, including
the $Z'$ contribution for  different $Z'$ masses as well as the one
expected for the SM, for the design value of $\sqrt{s}$ of $14$ TeV 
as well as its current value of  $7$ TeV.
\FIGURE[!ht]{
\epsfig{file=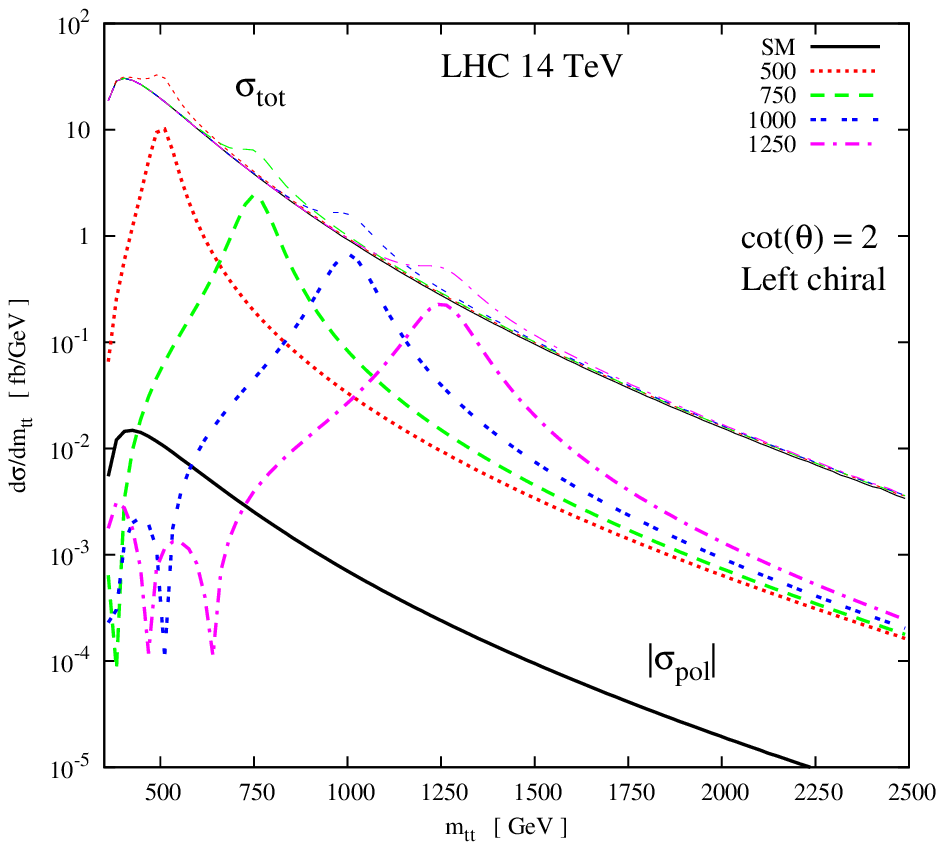,width=7.40cm}
\epsfig{file=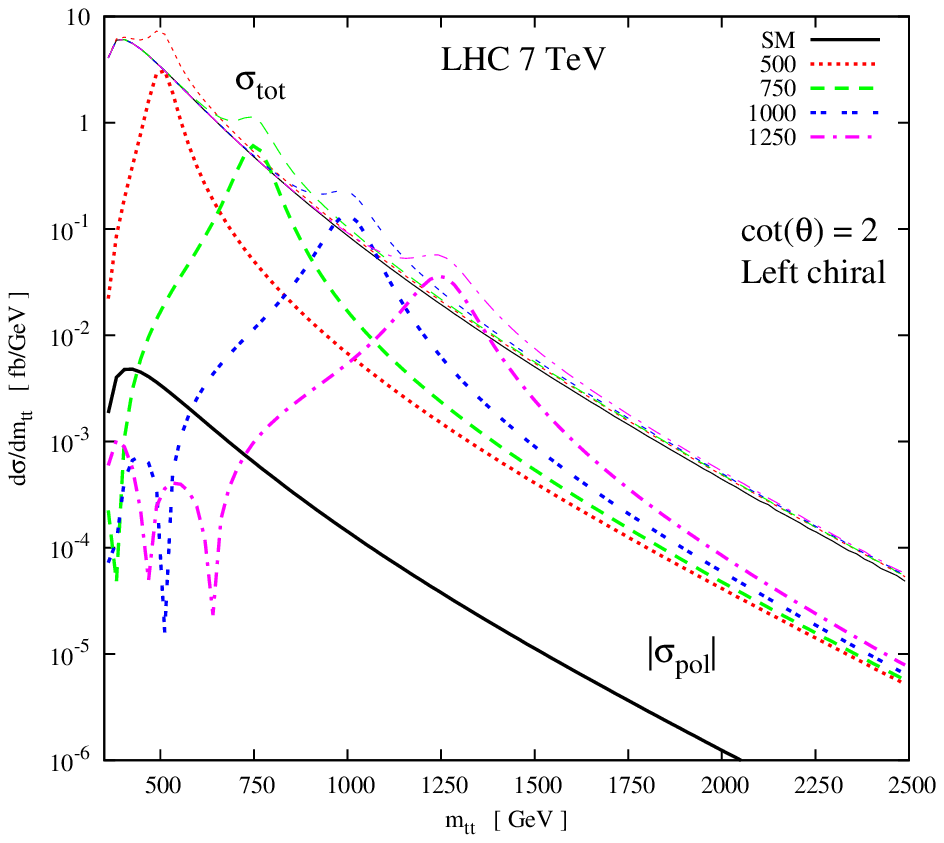,width=7.40cm}
\caption{\label{fig:mttLHC}The $m_{t\bar t}$ distribution of total 
cross-section $\sigma_{\rm tot}$ (thin lines) and polarized part
$|\sigma_{\rm pol}|$ (thick lines) are shown for the SM (solid/black lines)
and with $Z'$ of mass $500$ GeV (small-dashed/red lines), $750$ GeV 
(long-dashed/green lines) $1000$ GeV (double-dashed/blue lines) and $1250$ GeV
(dot-dashed/magenta lines) at $14$ TeV LHC (left panel) and $7$ TeV LHC (right
panel). We have assumed $\cot(\theta) = 2$ and the left chiral couplings of 
$Z'$ as in the Little Higgs model. }}

Fig.~\ref{fig:mttLHC} shows that the distributions in the total cross-section 
peak at the
respective $Z'$ masses. Not only that, even the polarization dependent
part peaks at the respective $Z'$ masses, showing that the major
contribution to the polarization comes from the chiral $Z'$ coupling. 
On the other hand, the polarization dependent part of the cross section
for the SM is lower by about 3 orders of magnitude.
Since we calculate these distributions with
left chiral couplings of $Z'$ the polarized part is negative near the 
resonance, i.e. $\sigma(+,+) < \sigma(-,-)$.  For sake of convenience,
we plot the absolute value $|\sigma_{\rm pol}|$ in Fig.~\ref{fig:mttLHC}. For the 
right chiral couplings the distribution is almost identical to that for the 
left chiral case, but with $\sigma(+,+) > \sigma(-,-)$ and is hence not shown.
It is thus expected that at least in the region of the resonance, the top
polarization would be a good measure of the chirality of the couplings.

Fig.~\ref{fig:mttLHC} also shows certain other interesting 
features, which do not directly concern us here. For example,
sign changes in $\sigma_{\rm pol}$ arising when contributions with different
$s$ channel exchanges interfere show up as sharp dips in the
distribution.

We also see that for $\sqrt{s}=7$ TeV, $[\sigma(+,+)-\sigma(-,-)]_{Z'} > 
[\sigma(+,+)+\sigma(-,-)]_{SM}$ near the resonance, for $M_{Z'} \ge 750$ GeV.
This means that in this case, it will be easier to distinguish the presence 
of $Z'$ from the SM background than in the case of $\sqrt{s}=14$ TeV.
For the latter the increased and dominant $gg\to t\bar t$ contribution 
causes a reduction in the polarization of the top quark.

In Fig.~\ref{fig:PtMZLHC} we show the 
degree of top polarization, $P_t = \sigma_{\rm pol}/\sigma_{\rm tot}$ 
as a function of $M_{Z'}$ for different  values of
the coupling $\cot(\theta)$ for 
$\sqrt{s}=14$ and for $\sqrt{s}=7$ TeV. 
\FIGURE[!ht]{
\epsfig{file=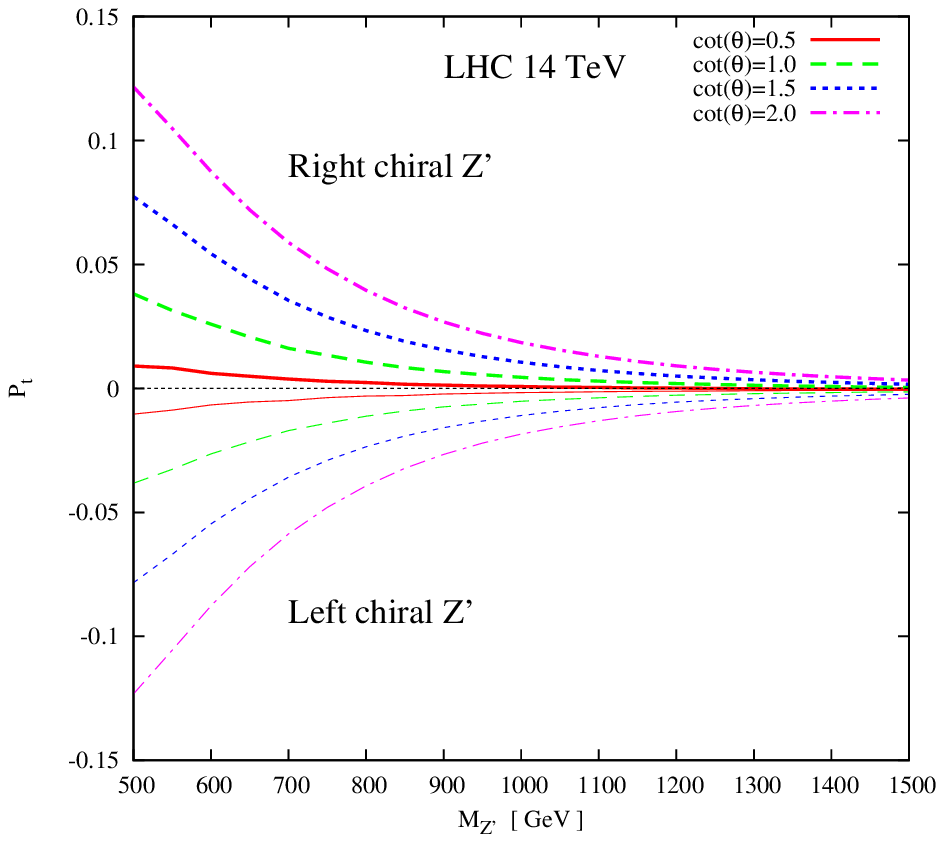,width=7.40cm}
\epsfig{file=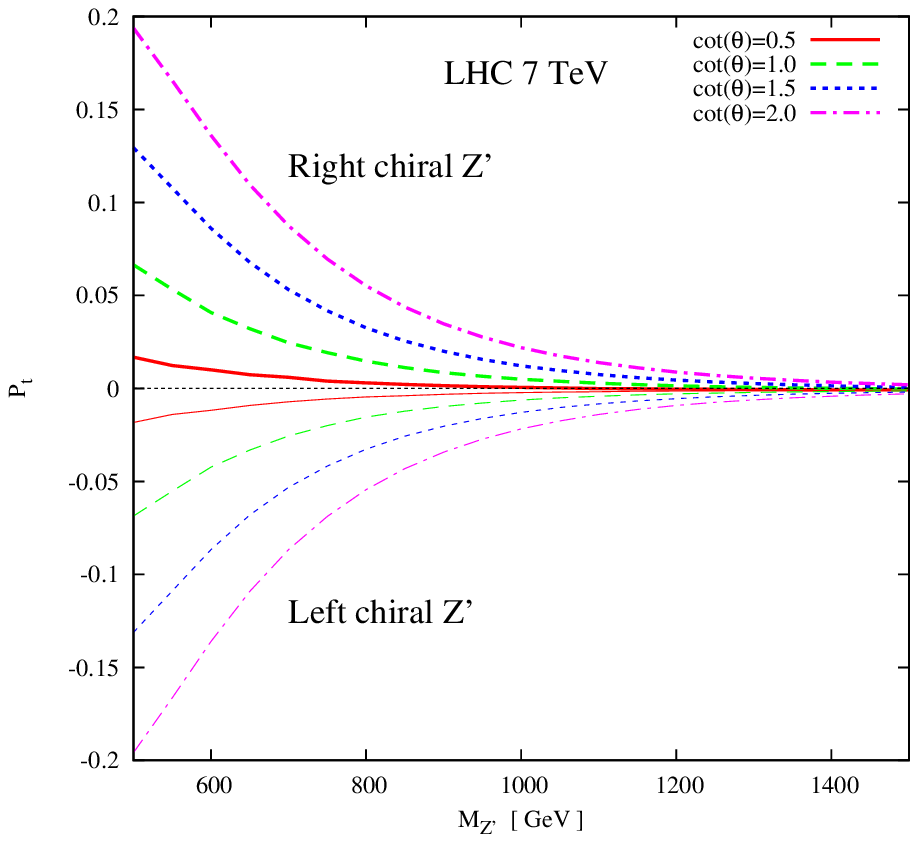,width=7.40cm}
\caption{\label{fig:PtMZLHC}The $M_{Z'}$ dependence of the top 
polarization $P_t$
for $\sqrt{s}=14$ TeV  (left panel) and
$\sqrt{s} = 7$ TeV LHC (right panel). The thick lines are
for the right-chiral coupling of $Z'$ and the thin lines are for the 
left-chiral couplings. The curves are shown for $\cot(\theta)=0.5$ (red/solid
line), $\cot(\theta)=1.0$ (green/big-dashed line), $\cot(\theta)=1.5$ 
(blue/small-dashed line) and $\cot(\theta)=2.0$ (magenta/dash-dotted line).}}
Since the SM contribution to the top polarization coming from the off-shell
$Z$ boson is very small, $|P_t^{SM}|<10^{-3}$ (see Fig.~\ref{fig:mttLHC}), we
have not shown it.  However, the full contribution from
interference terms involving $\gamma, \ Z$ and $Z'$ exchanges is taken into
account in all observables considered here. 
The relatively higher $q\bar q$ fluxes at the Tevatron, owing to it
being a $p\bar p$ collider, and rather small $gg$ flux because of its
lower energy, leads to rather large values of expected top polarizations
at the Tevatron, reaching 40\%.

The top polarization in the 
presence of $Z'$
is positive for  right chiral couplings of $Z'$ and negative for left 
chiral couplings.  This is because the dominant 
contribution to the polarization comes for $m_{t\bar t}$ near $Z'$ pole, where
the top polarization is dictated by the chirality of its couplings. 
This
suggests that a cut on  $m_{t\bar t}$, like $|m_{t\bar t}-M_{Z'}| \leq 
2\Gamma_{Z'}$ to select the $Z'$ pole, will increase the net polarization of
the top quark sample and also the sensitivity of any observable sensitive to
the top polarization.

Similarly, one can also look at the transverse momentum distribution of the 
top quark for signal enhancement. For a resonance of mass $M_{Z'}$ in the 
$t\bar t$ pair production, there is a peak at $m_{t\bar t} = M_{Z'}$ which
translates to a peak in the transverse momentum at $p_t^T = \beta(M_{Z'}^2) 
M_{Z'}/2$, where $\beta(s) = \sqrt{1-4m_t^2/s}$.
\FIGURE[!ht]{
\epsfig{file=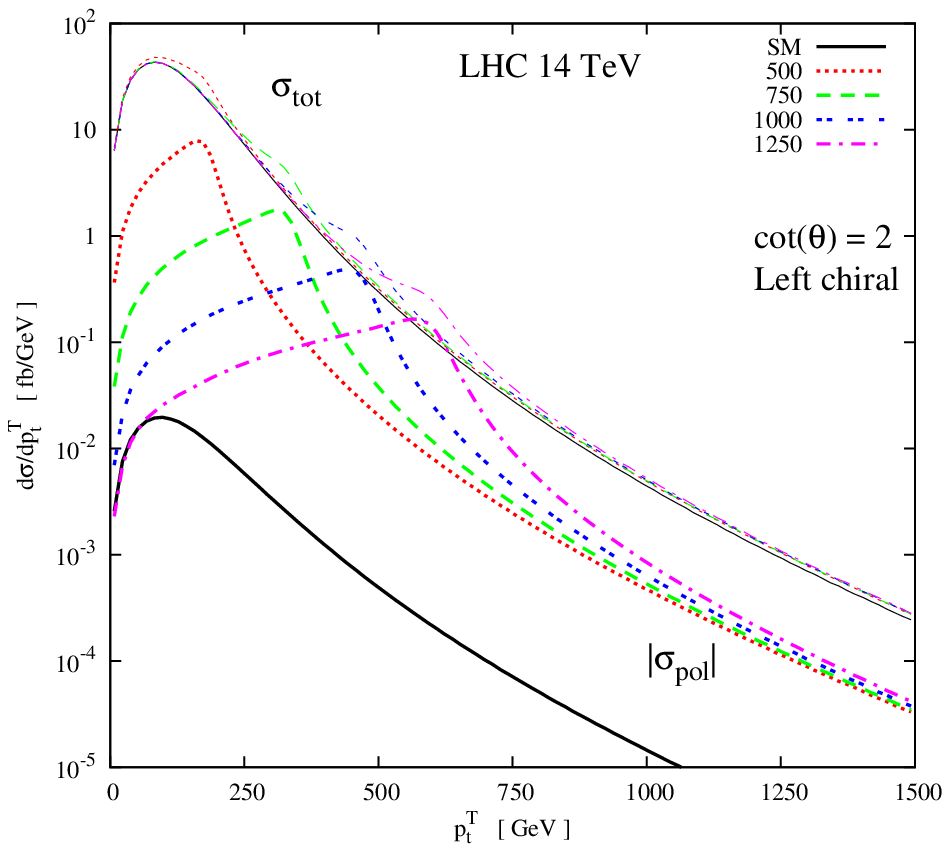,width=7.40cm}
\epsfig{file=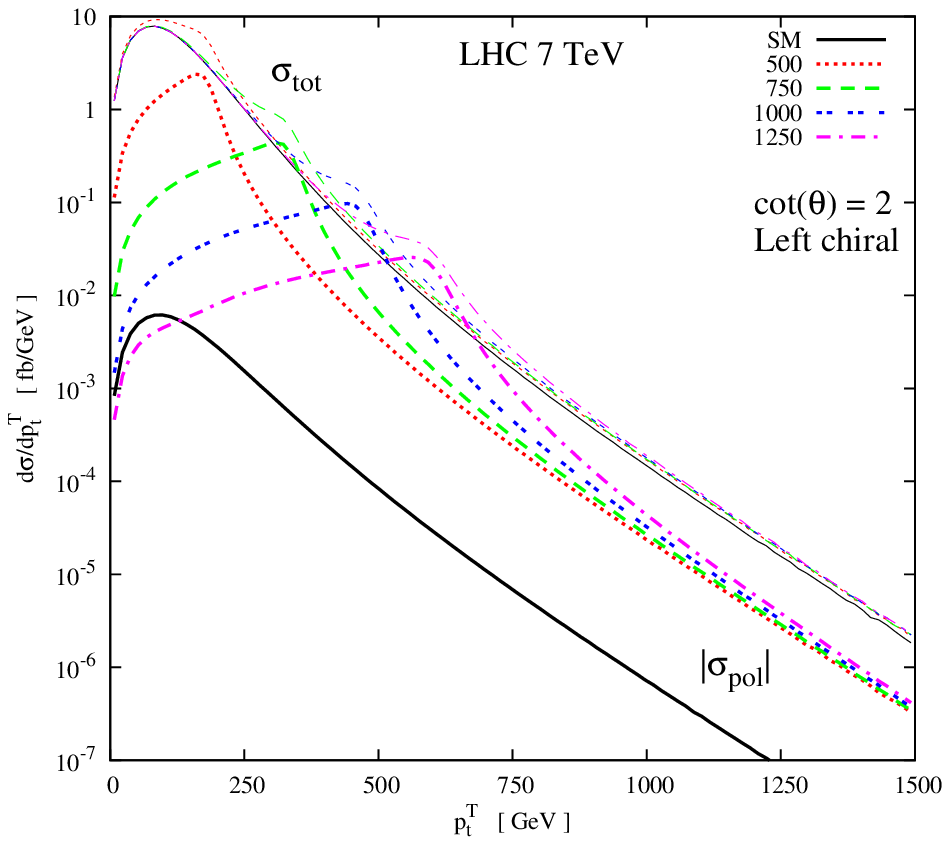, width=7.40cm}
\caption{\label{fig:ptdist}The $p_t^T$ distribution for the unpolarized 
cross-section
$\sigma_{\rm tot}$ (thin lines) and for the polarization dependent part
$|\sigma_{\rm pol}|$ (thick lines) for $\sqrt{s}=14$ TeV 
(left panel) and $\sqrt{s} = 7$ TeV (right panel). The peak in the distribution occurs
at $p_t^T = \beta_M M_{Z'}/2$ where $\beta_M=\sqrt{1-4m_t^2/M_{Z'}^2}$.
The legend is the same as in Fig.~\ref{fig:mttLHC}.} }
This peak is shown in the $p_t^T$ distribution of total cross-section 
$\sigma_{\rm tot}$ and polarized part $|\sigma_{\rm pol}|$ in Fig.~\ref{fig:ptdist}.
One can thus put a cut on the transverse momentum to improve the sensitivity 
of the polarization observables. The $p_t^T$ cut will turn out to be useful,
as we will see in the following sections.

We now study the use of the azimuthal distribution of the charged lepton
coming from top decay as a tool to measure the top polarization.
 
\subsection{Lepton azimuthal distribution}
 
To define the azimuthal angle of the decay products of the top quark we
choose the proton beam direction as the $z$ axis and
the top production plane as the $x-z$ plane, with top direction chosen to
have positive $x$ component. 
At the LHC, since the initial state has identical particles, 
the $z$-axis can point in the direction of either proton. This symmetry
implies that one cannot distinguish between an azimuthal angle $\phi$ and
an angle $2\pi- \phi$. 

In Fig.~\ref{fig:phiLR} we compare the normalized distributions  of the 
azimuthal angle $\phi_{\ell}$ 
of the decay leptons calculated using eq. (\ref{dsigell}) for three
cases, viz.,  (i) when the top quark has
negligible polarization, $|P_t|\approx 10^{-3}$, 
as in the SM (black/solid line),
 (ii) when the top has 100\% right-handed polarization, calculated
keeping only the $\sigma(+,+)$ in 
eq.(\ref{dsigell}) (green/big-dashed line) and (iii) when the top has
100\% left-handed polarization, calculated keeping only the $\sigma(-,-)$ in
eq.(\ref{dsigell}) (blue/small-dashed line).
As compared to the distribution for the (almost) unpolarized top
in the SM, a positively polarized top leads to a
distribution that is more sharply peaked near $\phi_\ell=0$.
The behaviour for a negatively polarized top is opposite, and the
relative number of leptons near $\phi_\ell=0$ is far reduced. 
Thus, it is clear from the 
Fig.~\ref{fig:phiLR} that the azimuthal distribution can easily distinguish 
between $100\%$ positive and $100\%$ negative top polarizations. 
However, in practice, the produced top has partial polarization described
by simultaneously non-zero values of $\sigma(+,+)$ as well as $\sigma(-,-)$
and also the spin-coherence contributions coming from the 
off-diagonal terms $\sigma(\pm,\mp)$. 
\FIGURE[!ht]{
\epsfig{file=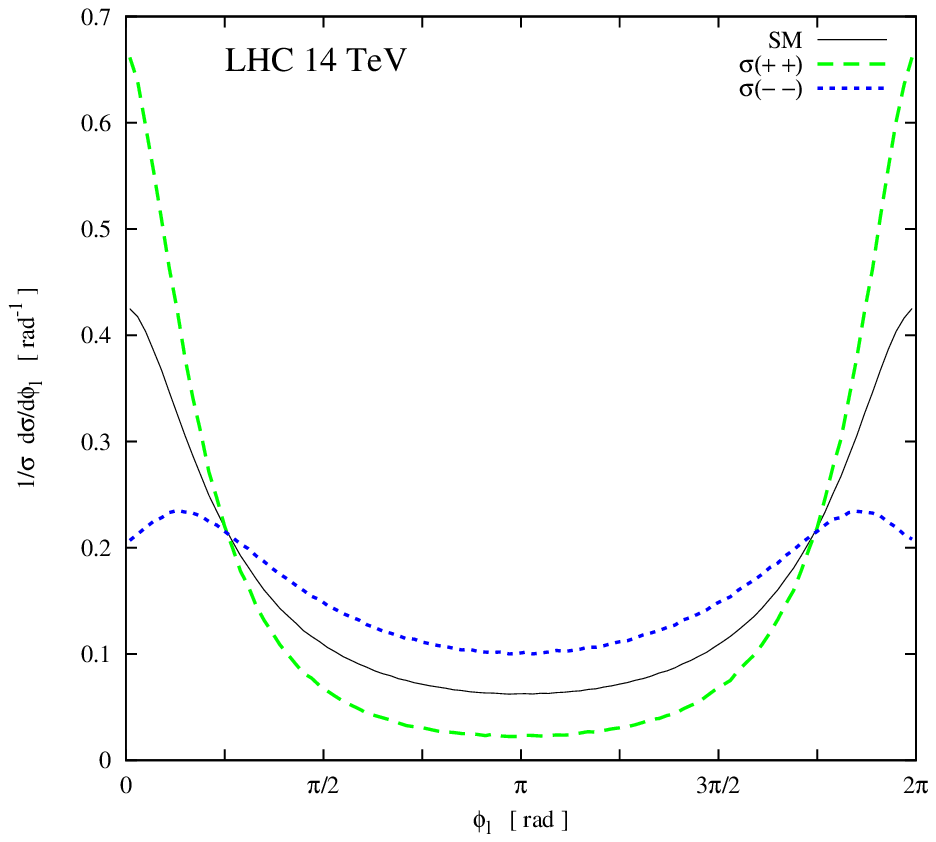, width=9.40cm}
\caption{\label{fig:phiLR}The normalized $\phi_{\ell}$ distribution of the decay
lepton for the SM: full contribution is shown in black/solid line, the
positive helicity contribution from $\sigma(+,+)$ is shown in green/big-dashed
line and the negative helicity contribution from $\sigma(-,-)$ is shown in
blue/small-dashed line. 
} }

The actual $\phi_{\ell}$ distributions for $M_{Z'}=500$ and $750$ GeV with
left and right chiral couplings, together with the SM distribution, 
are shown in  
Fig.~\ref{fig:phil_full}. The figure clearly
shows that for $M_{Z'}=500$ GeV with right chiral couplings, which yields 
positive top polarization, there is 
greater peaking of the leptons
near $\phi_\ell=0$ as compared to the unpolarized SM distribution. 
Similarly, for
left chiral couplings of the $Z'$ corresponding to a negatively polarized top 
sample, the peak 
near $\phi_\ell=0$ is reduced. In other words the 
qualitative behaviour of the $\phi_{\ell}$ distribution for $M_{Z'}=500$ GeV is 
same as for the completely polarized top quarks. This, however, is not the 
case for $M_{Z'}=750$ GeV as can be seen from the right panel of 
Fig.~\ref{fig:phil_full}. 

\FIGURE[!ht]{
\epsfig{file=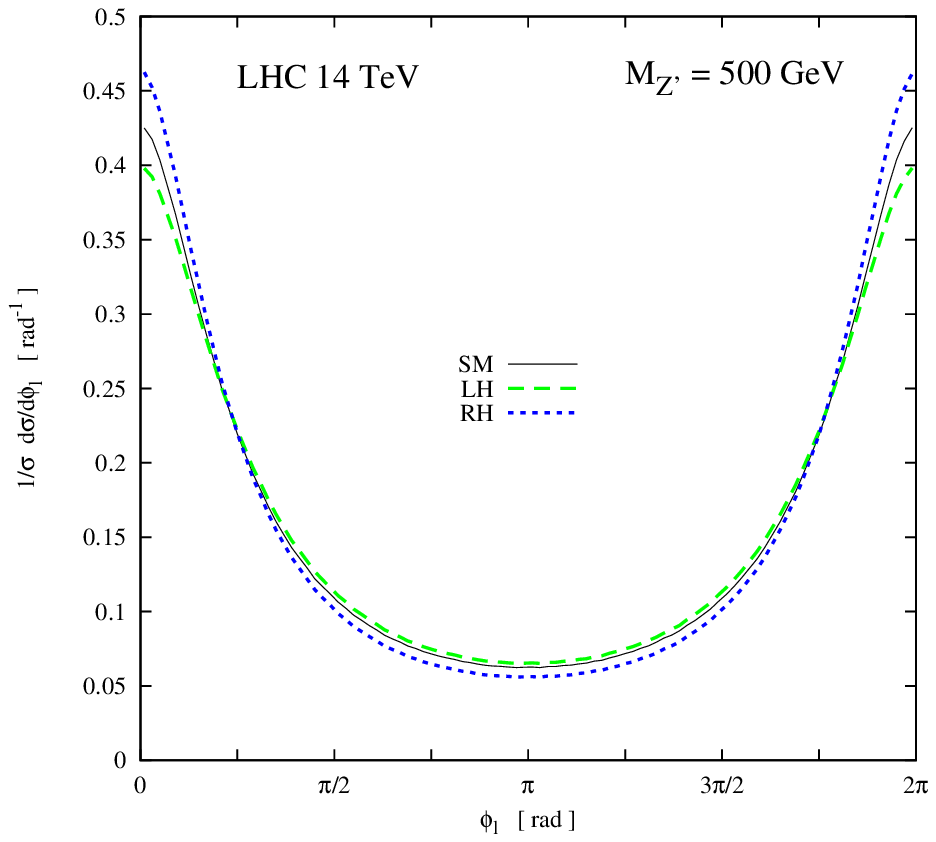, width=7.40cm}
\epsfig{file=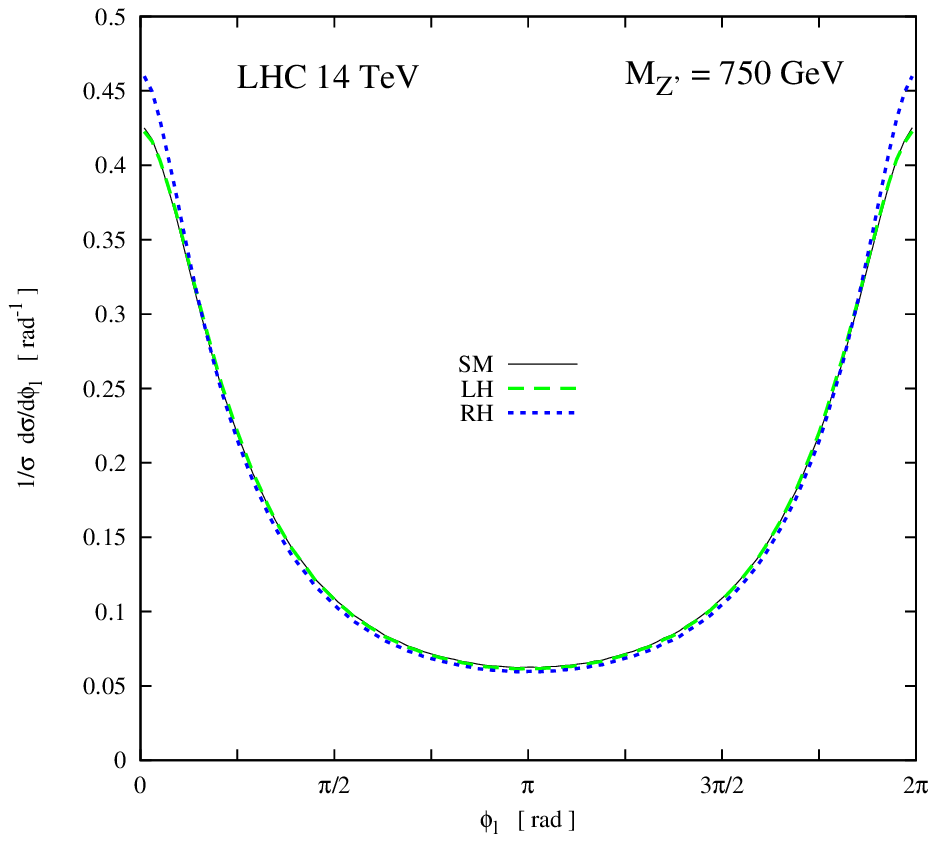, width=7.40cm}
\caption{\label{fig:phil_full}The $\phi_{\ell}$ distribution of the decay lepton for 
the SM (black/solid line), $Z'$ with left chiral (green/big-dashed line) and
right chiral (blue/small-dashed line) couplings. No kinematical cut has been
applied.} }

\def \ctl{$\cos\theta_{t\ell}$}
To understand this observed change in the  polarization dependence of the 
$\phi_{\ell}$ distribution as we go from  $M_{Z'} = 500$ GeV to a higher value
of $750$ GeV (as well as to understand its dependence  on the top transverse 
momentum) we need to
see how the $\phi_{\ell}$ distribution in the laboratory frame is related to
the simple angular distribution given in eq. (\ref{topdecaywidth}) with
$\kappa_f = 1$. The corresponding distribution in the laboratory
frame, on using the relation 
\beq\label{boost}
\cos\theta_{\ell}^* = \frac{\cos\theta_{t\ell} - \beta}{1 - \beta
\cos\theta_{t\ell}}
\eeq
between the angle $\theta_{\ell}^*$ between the top spin and the lepton
direction in the rest frame of the top and the angle $\theta_{t\ell}$
between the top and lepton directions in the laboratory
frame, becomes
\beq\label{thetatldist}  \displaystyle
\frac{1}{\Gamma_{\ell}}\frac{d\Gamma_{\ell}}{d\cos\theta_{t\ell}} = \displaystyle \frac{1}{2}
(1-\beta^2)(1 - P_t \beta)\frac{1 + \frac{P_t - \beta}{1 - P_t
\beta} \cos\theta_{t\ell}}{(1- \beta
\cos\theta_{t\ell})^3},
\eeq
where $\beta = \sqrt{1 - m_t^2/E_t^2}$, and 
\beq\label{costhetatl}
\cos\theta_{t\ell} = \cos\theta_t \cos\theta_{\ell} + \sin\theta_t \sin\theta_{\ell}
\cos\phi_{\ell}.
\eeq
In practice, the distribution would also have to be integrated over
$\theta_t$, $\theta_{\ell}$ and the lepton energy. 

The first thing to note about the distribution of (\ref{thetatldist}) is
that because of the denominator, there is peaking for large \ctl, and
hence for small $\phi_{\ell}$, according to eq. (\ref{costhetatl}). Thus, the
boost produces a collimating effect along the direction of the top
momentum, which gets translated to a peaking at $\phi_{\ell} = 0$. Secondly,
unless $P_t = \pm 1$, the form of the distribution
depends on relative values of $P_t$ and  typical values of $\beta$, in
the combination
\beq
P_t^{\rm eff} = \frac{P_t - \beta}{1 - P_t \beta}.  
\eeq
Thus there is a polarization dependent effect and an effect which occurs
simply because of the boost, independent of the polarization, which could
compete with each other.
Thus, the dependence on $\phi_{\ell}$ would turn out
to be controlled by $P_t$ so long as typical values of $\beta$ are small
compared to $P_t$. This condition would be satisfied for smaller values
of $E_t$ and hence of $M_{Z'}$, since the major contribution comes from
$m_{t\bar t} \approx M_{Z'}$. 
This helps us to understand why, as mentioned earlier, the behaviour of
the $\phi_l$ distribution is the same as that of completely polarized
tops for $M_{Z'} = 500$ GeV, but not for $M_{Z'} = 750$ GeV.
As we shall see, this will be useful in devising a suitable cut.

In the above reasoning, in order to illustrate the major effects of the
change of frame from the top rest frame to the laboratory frame, 
we have taken a simplified approach,
characterizing all top spin effects in terms of the longitudinal
polarization $P_t$. In all our calculations, however, we deal with the full 
spin density matrix of the top quark, using eq.~(\ref{matelsq}).  

To characterize the shape of the $\phi_{\ell}$ distribution, it is convenient
to define a lepton azimuthal asymmetry
\cite{usleshouch,Djouadi:2007eg,Godbole:2009dp}:
\begin{equation}
A_{\ell} = \frac{\sigma(\cos\phi_{\ell} >0) - \sigma(\cos\phi_{\ell} <0)}
{\sigma(\cos\phi_{\ell} >0) + \sigma(\cos\phi_{\ell} <0)} =
\frac{\sigma(\cos\phi_{\ell} >0) - \sigma(\cos\phi_{\ell} <0)}{\sigma_{\rm tot}} \ \ .
\label{eq:Al}
\end{equation}
This asymmetry is non-zero for the SM. While it is expected to be
substantially different from the SM value of $\sim 0.52$, 
when right chiral couplings of
$Z'$ are included,  Fig.~\ref{fig:phil_full} indicates that for larger
values of $M_{Z'}$, the asymmetry for left chiral couplings may not be very 
different from that for the SM.

In Fig.~\ref{fig:al_full_mtt50}, left panel, we show the deviation $\delta
A_{\ell}$ of the lepton azimuthal asymmetry from the SM value as a function of
$M_{Z'}$ for different values of right and left chiral couplings. 
We see that while $\delta A_{\ell}$ can characterize well the
polarization for the case of right
chiral $Z'$ couplings, it can discriminate left chiral couplings only for
$M_{Z'}$ values below about 600 GeV.
Ideally, we would like 
$\delta A_{\ell}$ to be a monotonic
function of the top polarization for some choice of kinematics. 
In what follows,
we investigate the possibility of finding suitable kinematical cuts, which when applied, makes 
$\delta A_{\ell} 
= A_{\ell} - A_{\ell}^{SM}$ a monotonically increasing function of the top
polarization irrespective of the mass of $Z'$.

\subsection{Kinematic cuts for $A_{\ell}$}
As can be seen from Figs.~\ref{fig:mttLHC} and \ref{fig:ptdist}, the $Z'$
resonance contributes to the unpolarized cross-section $\sigma_{\rm tot}$ 
and polarization
$\sigma_{\rm pol}$ in the same kinematic region. In other words, there is a peak
in the $m_{t\bar t}$ distribution at $m_{t\bar t}^{pole}=M_{Z'}$ and a peak in 
the $p_t^T$ distribution at $(p_t^T)^{peak}=\beta(M_{Z'}^2)M_{Z'}/2$ for both
$\sigma_{\rm tot}$ and $\sigma_{\rm pol}$. 
Thus a cut on $m_{t\bar t}$
and/or $p_t^T$ will help select the events with large top polarization and
hence large contribution to $A_{\ell}$. 
We study below the effects of these cuts on the shape of the $\phi_\ell$
distribution and on the lepton asymmetry $A_\ell$.

Since the $Z'$ resonance appears at $m_{t\bar t}=M_{Z'}$, it is simple to 
imagine the importance of the $m_{t\bar t}$ cut. The $p_t^T$ cut can be 
motivated as follows:
As seen in Fig.~\ref{fig:phil_full} (right panel), the $\phi_{\ell}$ distribution
and hence the lepton asymmetry $A_{\ell}$ for the left chiral $Z'$ is almost same
as that of the SM. This happens because the left chiral $Z'$ produces 
negatively polarized top quarks with high transverse momentum. Here the 
negative polarization tends to diminish the peaking near $\phi_{\ell}=0$ of 
the leptons but the large transverse 
momentum of these highly polarized tops provides a larger factor of
$\sin\theta_t$ in front of $\cos\phi_{\ell}$ in eq. (\ref{costhetatl}), increasing
the peaking, and the two effects can cancel each other. 
The kinematic effect always leads
to collimation hence it only adds to the effect generated by positively 
polarized tops (right chiral $Z'$), while for negative top polarization it 
reduces the effect if not cancel it. 
If the transverse momentum of the top quark is too large then the kinematic
effect may even over-compensate the de-collimating effect of the negative top
quark polarization.  On the other hand, for the process under consideration
the degree of polarization expected increases  with a lower cut on $p_t^T$.
Thus we need to choose a window of $p_t^T$ values such that the contribution
from $Z'$ is maximized and $\delta A_\ell$ reflects the sign of the
polarization

We have examined the effect of  the following three kinds of 
kinematic cuts:
\begin{itemize}
\item {\bf Resonant $m_{t\bar t}$ cut}: We select events with 
$m_{t\bar t}$ near $M_{Z'}$, i.e., $|m_{t\bar t}-M_{Z'}|<50$ GeV for each 
value of $M_{Z'}$ and $\cot(\theta)$. 
\item {\bf Fixed $p_t^T$ cut}: We select events in a fixed range of
the $p_t^T$ for each value of $M_{Z'}$ and $\cot(\theta)$. Examples of this
class of cuts chosen are $p_t^T > (p_t^T)^{min}$, where $(p_t^T)^{min} = 300,
\ 400$ or $500$ GeV.
\item {\bf Adaptive $p_t^T$ cut}: We select events in a range of
transverse momentum where the lower and the upper cuts depend both on the
$M_{Z'}$ and $\cot(\theta)$ (via the decay width). One such cut is
$p_t^T \in [ \ \beta(M_{Z'}^2)(M_{Z'}-2\Gamma_{Z'})/2, \ \ 
\beta(M_{Z'}^2)(M_{Z'}+2\Gamma_{Z'})/2 \ ]$.
\end{itemize}
We show $\delta A_{\ell}$ as a 
function of $M_{Z'}$ and various values of $\cot(\theta)$, for both left and
right chiral couplings, and without any kinematic cut, in the left panel of 
Fig.~\ref{fig:al_full_mtt50}.
\FIGURE[!ht]{
\epsfig{file=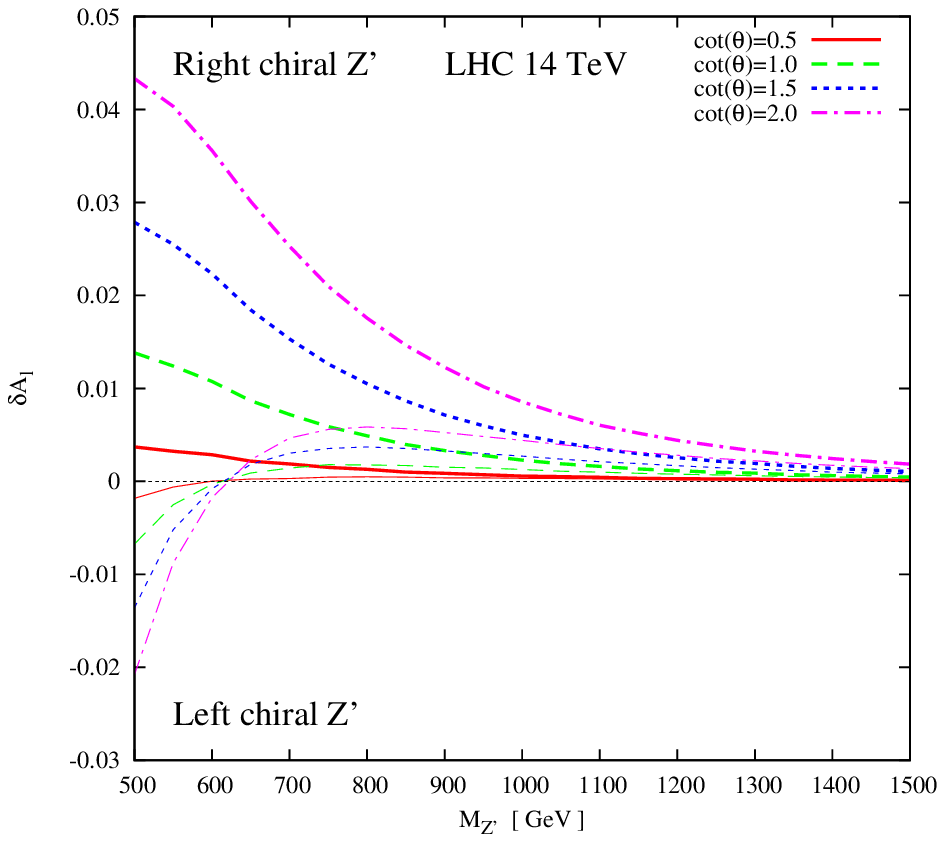,  width=7.40cm}
\epsfig{file=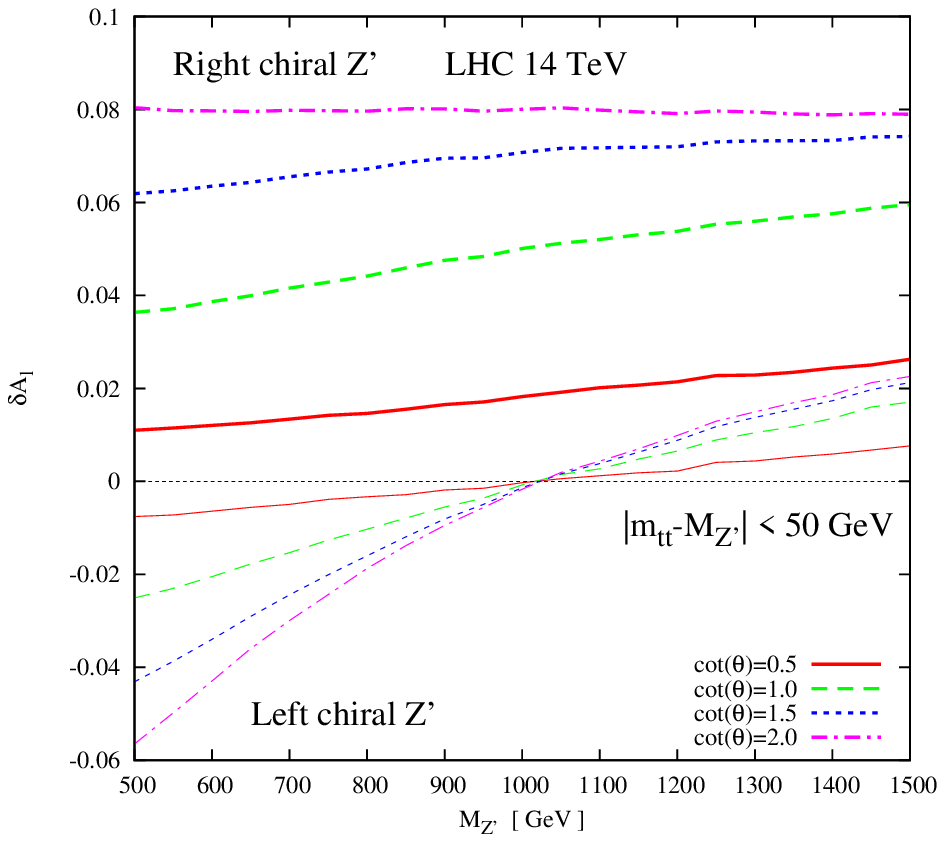, width=7.40cm}
\caption{\label{fig:al_full_mtt50}The $\delta A_{\ell}$ as a function of $M_{Z'}$ 
for $\cot(\theta)=0.5$ (red/solid line), $\cot(\theta)=1.0$ (green/big-dashed
line), $\cot(\theta)=1.5$ (blue/small-dashed line) and $\cot(\theta)=2.0$ 
(magenta/dash-dotted line). The thick lines are for right chiral couplings
and the thin lines are for left chiral couplings. The left panel is without
any kinematical cut while for the right panel a cut $|m_{t\bar t}-M_{Z'}|<50$
GeV is applied to enhance the $Z'$ resonance effect. } }
As mentioned earlier, for $M_{Z'}<600$ GeV the sign of $\delta A_{\ell}$ 
follows the chirality of the 
couplings and hence the sign of the top polarization. However, for
larger masses and left chiral couplings of $Z'$, $\delta A_{\ell}$ changes sign
because of the increased number of top events with large $p_t^T$. This, as
mentioned earlier, over-compensates the de-collimation due to the negative 
polarization and leads to $\delta A_{\ell}$ having sign opposite to that of the 
polarization. Thus a measurement of $\delta A_{\ell}$ without any kinematic cut 
cannot determine even the sign of the top polarization, let alone its magnitude.

We first apply the resonant $m_{t\bar t}$ cut, i.e. $|m_{t\bar t}-M_{Z'}|<50$
GeV and show the consequent $\delta A_{\ell}$ as a function of $M_{Z'}$ 
and various values of $\cot(\theta)$ in the
right panel of Fig.~\ref{fig:al_full_mtt50} for both left and right chiral 
couplings. With a cut on the invariant mass near the resonance, the resultant
top polarization is almost independent of $M_{Z'}$ when the decay width 
$\Gamma_{Z'}$ is larger than the range of the  $m_{t\bar t}$ cut window,
i.e., for $\cot(\theta) \ge 1.5$.
The corresponding asymmetry too is almost independent of $M_{Z'}$ for the
right chiral couplings. For the left chiral couplings, the asymmetry not only
varies with $M_{Z'}$, it also changes its sign near $M_{Z'}=1025$ GeV. Again,
the change of sign is due to polarization independent collimation over 
compensating the de-collimation caused by negative polarization. Thus, even
with the resonant selection cut on $m_{t\bar t}$ one cannot determine even
the sign of the top polarization. The cut helps increase the net top 
polarization and thus the change of sign  of $\delta A_{\ell}$ takes place at a
higher value of the $M_{Z'}$ as compared to the previous case, shown in
Fig.~\ref{fig:al_full_mtt50}, left panel.

The next cut studied is the fixed $p_t^T$ cut. We apply three different 
cuts: $p_t^T > 300, \ 400$ and $500$ GeV and show $\delta A_{\ell}$ as a 
function of $M_{Z'}$ for different values of $\cot(\theta)$ in 
Fig.~\ref{fig:al_full_pt} top-left, top-right and bottom-left panels, 
respectively. We see that the sign of 
$\delta A_{\ell}$ follows the chirality of the $Z'$ couplings for all three 
$p_t^T$ cuts and also the curves for different values of $\cot(\theta)$ are
ordered according to the values. In other words, with fixed $p_t^T$ cuts 
$\delta A_{\ell}$ is a monotonically increasing function of $\cot(\theta)$ 
and hence that of top polarization. 

This is not so for the third cut, the adaptive $p_t^T$ cut  
shown in the bottom-right panel of Fig.~\ref{fig:al_full_pt}. 
In this case the width of the window depends upon $\cot(\theta)$ 
through $\Gamma_{Z'}$. Thus for different values of $\cot(\theta)$, the
amount of QCD $t \bar t$ production included in the denominator in the
calculation of asymmetry differs.
This explains why the curve for $\cot(\theta)=2.0$ crosses the
curve for $\cot(\theta)=1.5$ for left chiral $Z'$ with adaptive cuts.
But this mostly reflects the somewhat higher effective value of the 
polarization that is attained even for a lower value of $\cot (\theta)$. 
\FIGURE[!ht]{
\epsfig{file=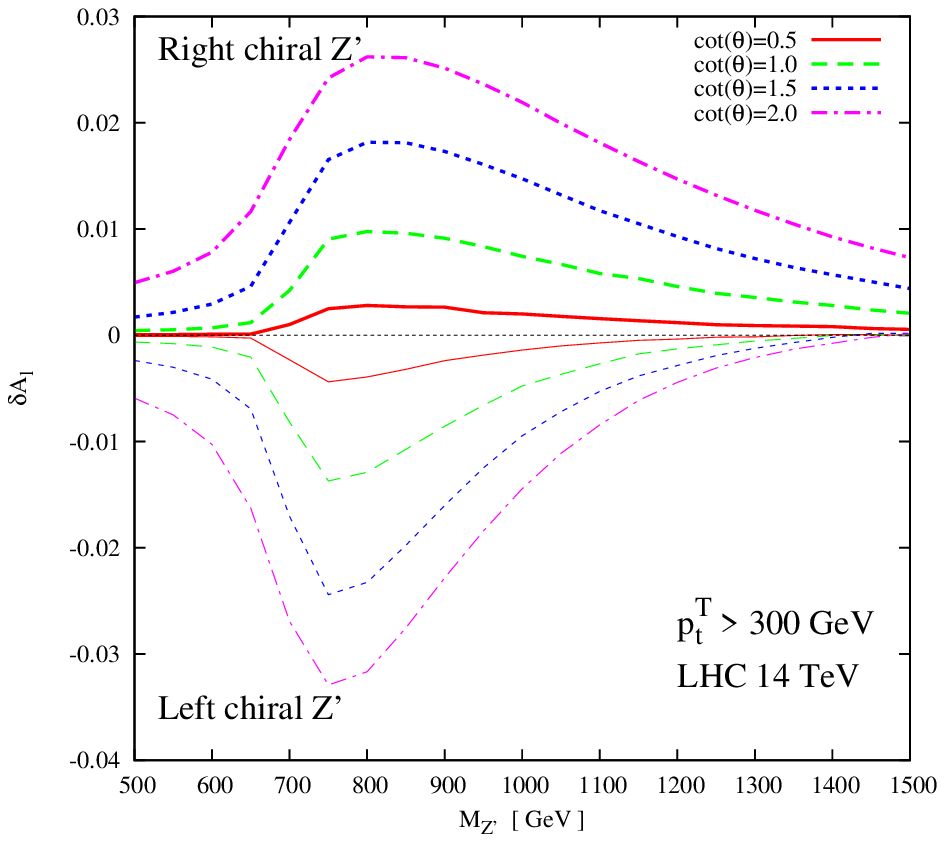, width=7.40cm}
\epsfig{file=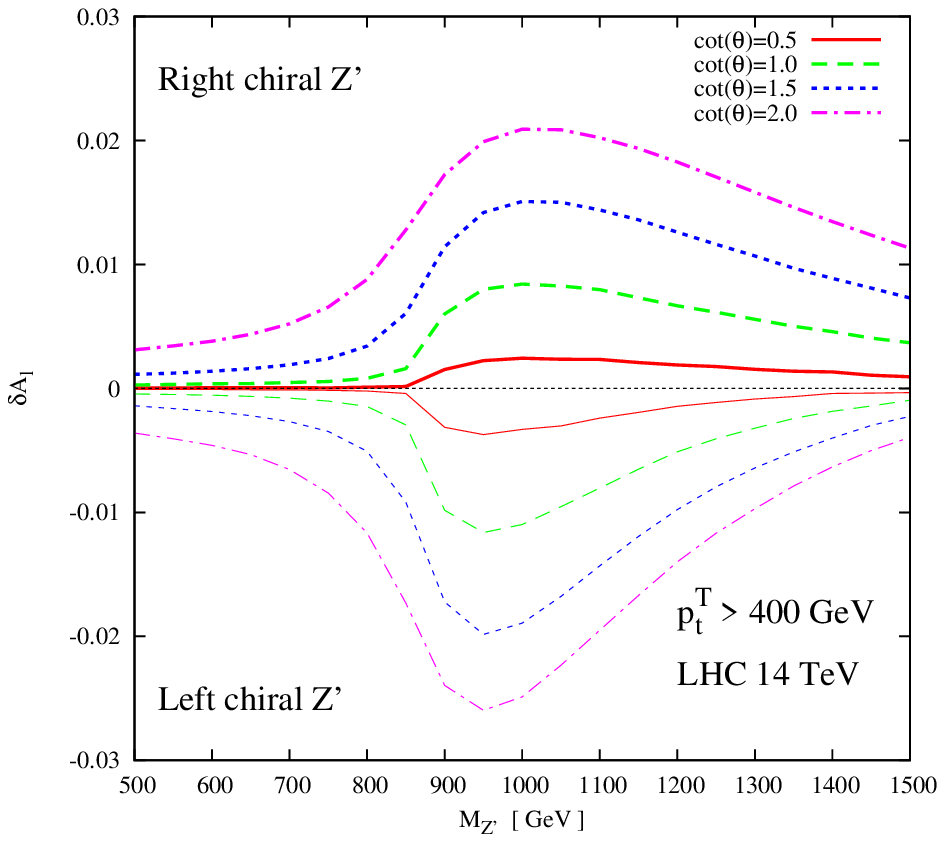, width=7.40cm}
\epsfig{file=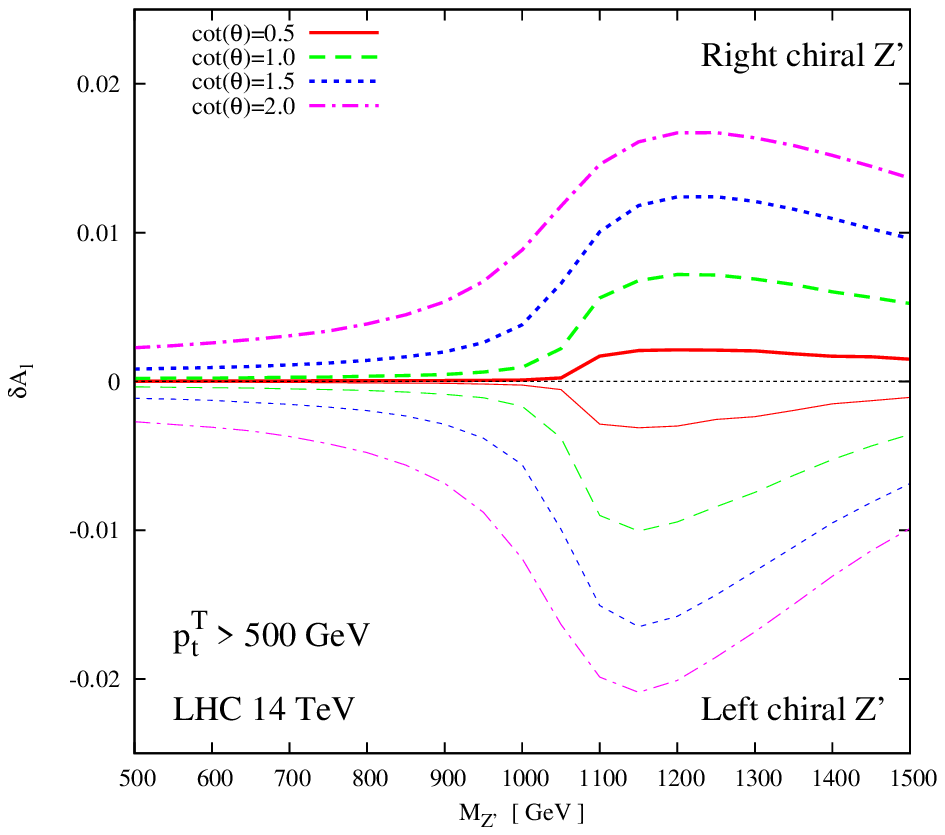, width=7.40cm}
\epsfig{file=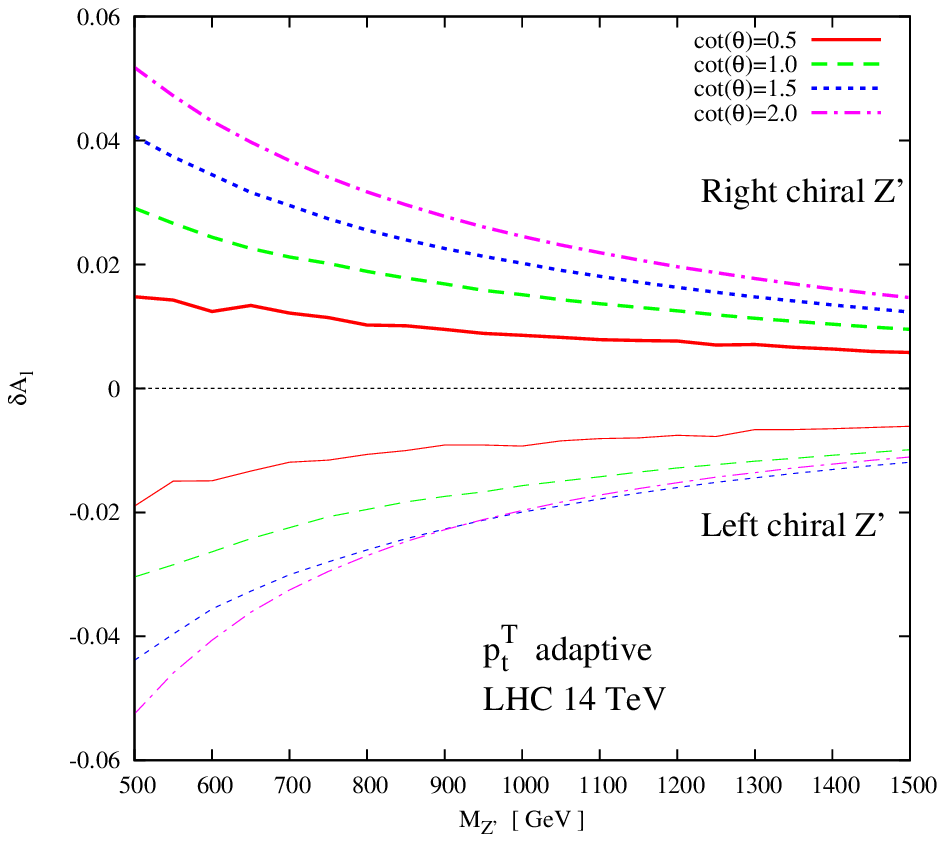, width=7.40cm}
\caption{\label{fig:al_full_pt}The $\delta A_{\ell}$ as a function of $M_{Z'}$ 
for $\cot(\theta)=0.5$ (red/solid line), $\cot(\theta)=1.0$ (green/big-dashed
line), $\cot(\theta)=1.5$ (blue/small-dashed line) and $\cot(\theta)=2.0$ 
(magenta/dash-dotted line). The thick lines are for right chiral couplings
and the thin lines are for left chiral couplings. The first three panels,
from left to right and top to bottom, correspond to different values of the 
cut on $p_t^T$  and the fourth panel corresponds to an adaptive $p_t^T$ cut
as described in the text.}}

As for the shapes of the $\delta A_{\ell}$ curves for fixed $p_t^T$ cuts, we note
that there is a peak in the $p_t^T$ distribution for $(p_t^T)^{peak} =
\beta(M_{Z'}^2)M_{Z'}/2$. For the $p_t^T>300$ GeV cut 
(Fig.~\ref{fig:al_full_pt} top-left panel) the peak of the $p_t^T$ distribution
is removed for $M_{Z'} < 695$ GeV. 
The rise in $|\delta A_{\ell}|$ starts when the transverse
momentum corresponding to $m_{t\bar t} = M_{Z'} + \Gamma_{Z'}$ appears above
the cut, i.e. when we have
$\beta((M_{Z'}+\Gamma_{Z'})^2)( M_{Z'}+\Gamma_{Z'})/2 > (p_t^T)^{min}$. 

To summarize, we have motivated and numerically demonstrated the 
monotonic behaviour of $\delta A_{\ell}$ as a function of the longitudinal top 
polarization
under fixed $p_t^T$ cuts. In case of the adaptive $p_t^T$ cut this is not 
as clear cut. 
We can say however that $p_t^T>300$ GeV and the adaptive $p_t^T$ cut
are two suitable choices for studying the top polarizations. We shall use
only these two cuts for other collider options in the later sections.

It is relevant to note here that a cut with $p_t^T>400$
GeV or higher makes the decay product of the top quarks highly collimated
and it may be difficult to extract the $\phi_{\ell}$ distribution from such  
collimated final states. The problem in the case of adaptive $p_t^T$ cut is 
more severe as the lower limit rises almost linearly with $M_{Z'}$. However, 
it has been reported \cite{MarcelVos} recently that reconstruction of the 
electron inside  the fat top jet is possible up to $p_t^T \sim 1000$ GeV, and 
it may be possible to extract the $\phi_{\ell}$ distribution with the help of 
sub-jet structure techniques even for heavy resonances with mass up to 2000 
GeV. Alternatively, one can use observables~\cite{Shelton:2008nq} constructed 
out of energies of the decay products  in the case of highly boosted 
tops~\cite{Krohn:2009wm}. However, it must be remembered that some of these 
observables will not be as  robust polarimeters as the lepton angular 
distribution, against the effect of anomalous couplings of the top.

\subsection{Statistical significance of $\delta A_{\ell}$}
We now study the statistical significance of  our chosen observable $\delta A_{\ell}$
under various conditions of $\sqrt{s}$ and luminosity. 
For this we define the sensitivity for the observable $\delta A_{\ell}$ 
as:
\begin{equation}
{\rm Sensitivity} (\delta A_{\ell}) = \frac{\delta A_{\ell}}{\Delta A_{\ell}^{SM}} =
\frac{ A_{\ell} - A_{\ell}^{SM}}{\Delta A_{\ell}^{SM}}, \hspace{1.0cm}\Delta A_{\ell}^{SM} =
\sqrt{\frac{1-(A_{\ell}^{SM})^2}{L \ \sigma_{\rm tot}}}
\end{equation}
\FIGURE[!ht]{
\epsfig{file=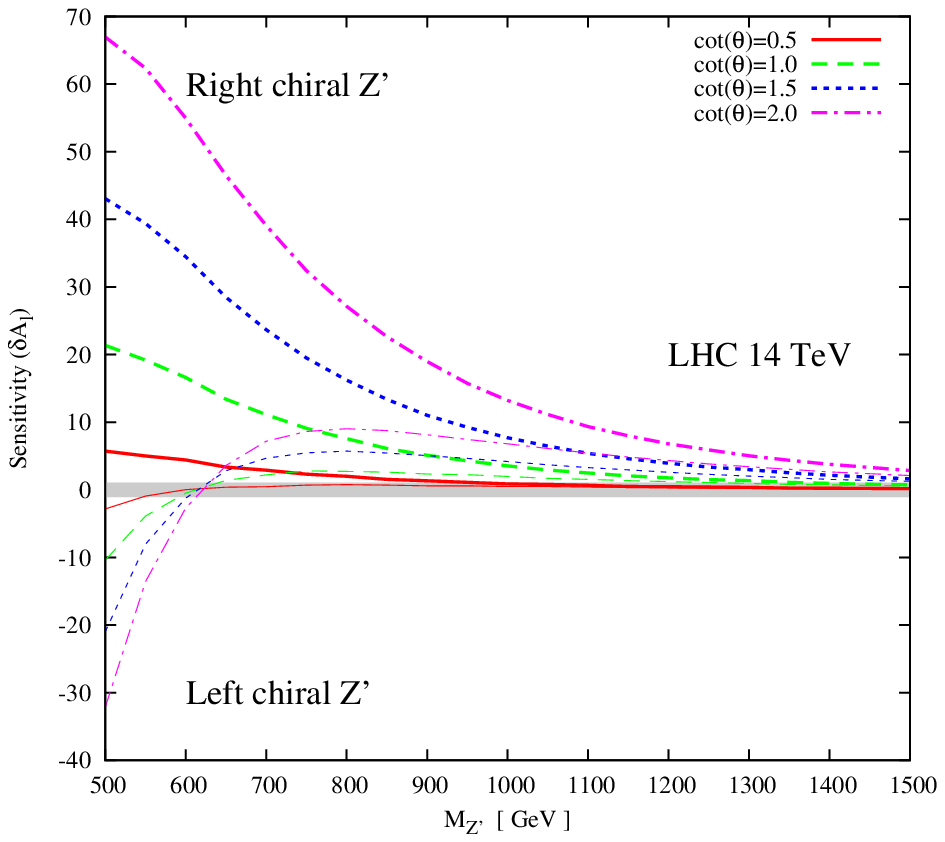,  width=7.40cm}
\epsfig{file=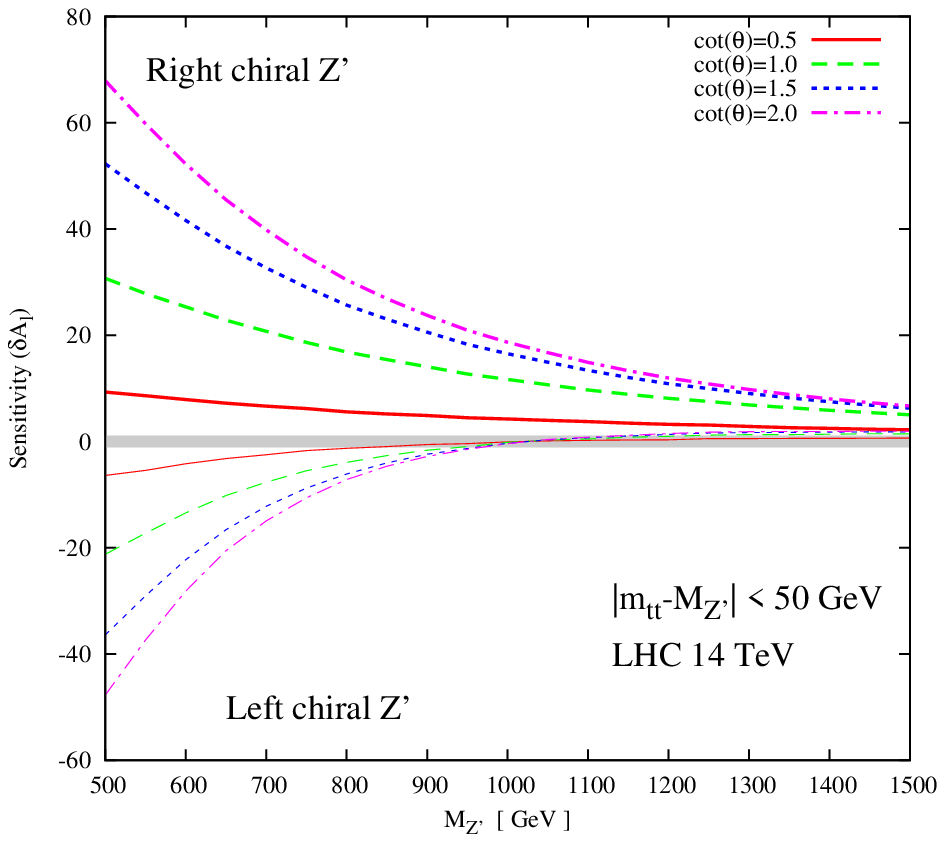, width=7.40cm}
\caption{\label{fig:sen_full_mtt50}The sensitivity for $\delta A_{\ell}$ as a 
function of $M_{Z'}$ for $\sqrt{s} = 14$ TeV and integrated luminosity of 
$10$ fb$^{-1}$ ($\ell =e,\mu$). 
The shaded region corresponds to sensitivity between $-1$ and $+1$.
The legend is the same as in Fig.~\ref{fig:al_full_mtt50}.
} }
The 
sensitivity defined above can be either positive or negative, depending
on the sign of $\delta A_{\ell}$. If the value of the sensitivity lies 
between $-1$ and $+1$, the corresponding $\delta A_{\ell}$ 
would not be distinguishable at the $1 \sigma$ level
from the SM prediction. 
This region is shown shaded in the sensitivity plots. We show the
sensitivities for $\sqrt{s} = 14$ TeV  and an integrated luminosity of 
$10$ fb$^{-1}$ for the cases of no kinematic 
cut and the simple $m_{t\bar t}$ cut in Fig.~\ref{fig:sen_full_mtt50}. These
sensitivities correspond to the asymmetries shown in 
Fig.~\ref{fig:al_full_mtt50}. 
\FIGURE[!ht]{
\epsfig{file=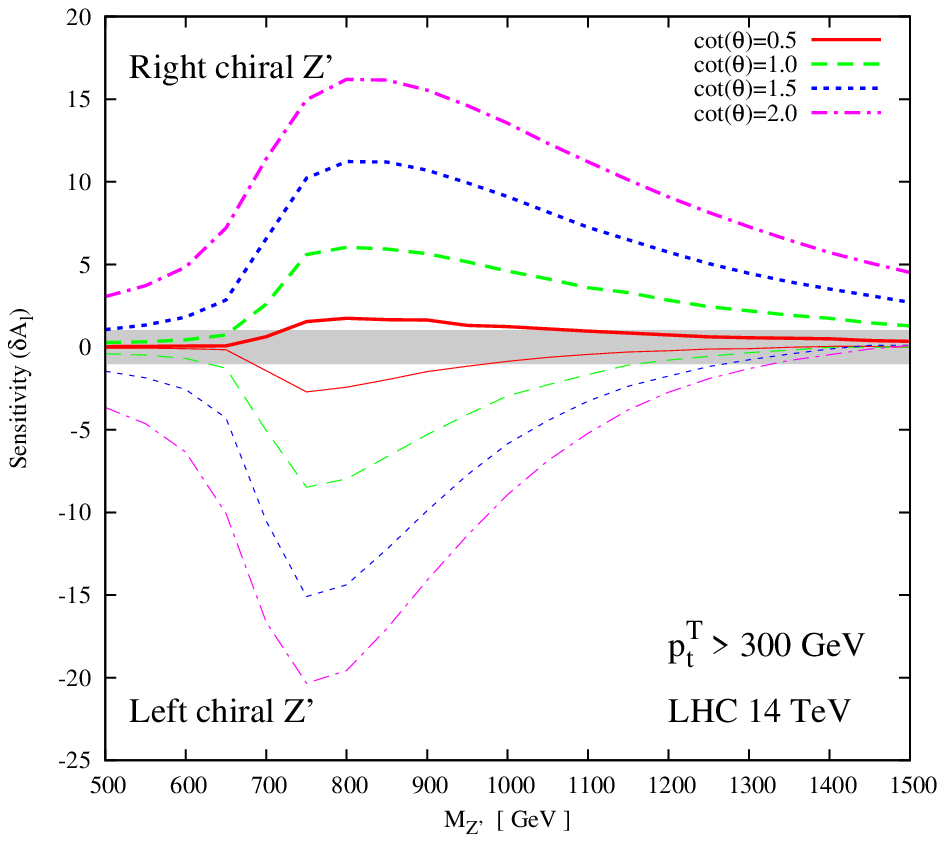, width=7.40cm}
\epsfig{file=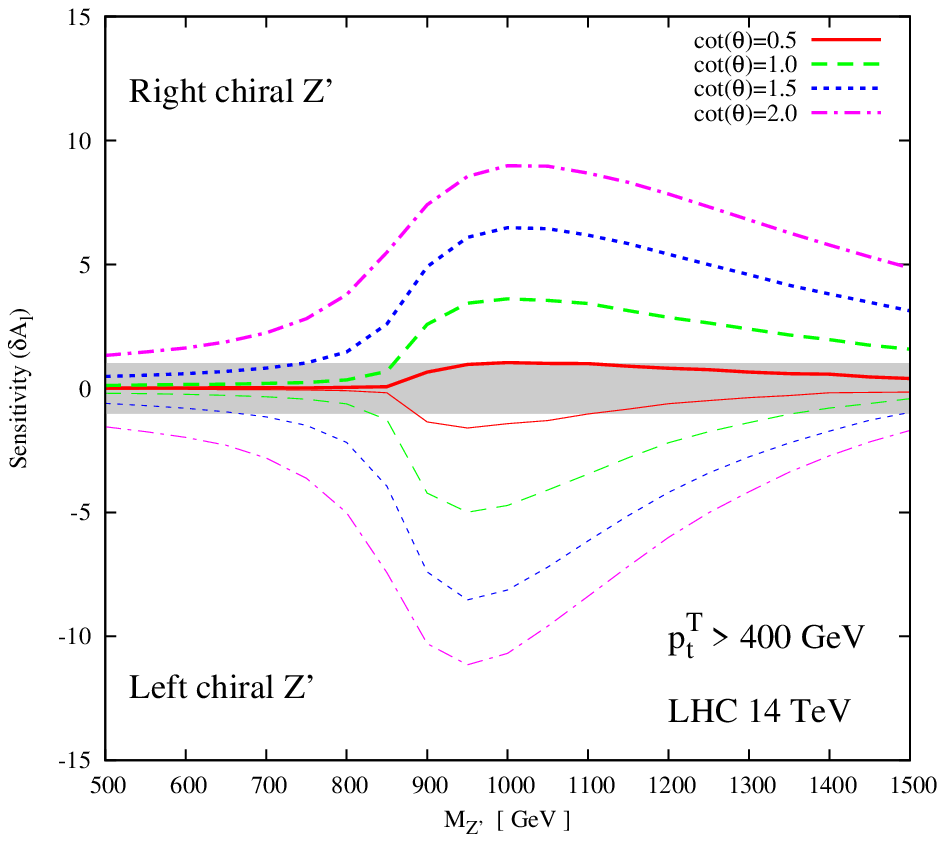, width=7.40cm}
\epsfig{file=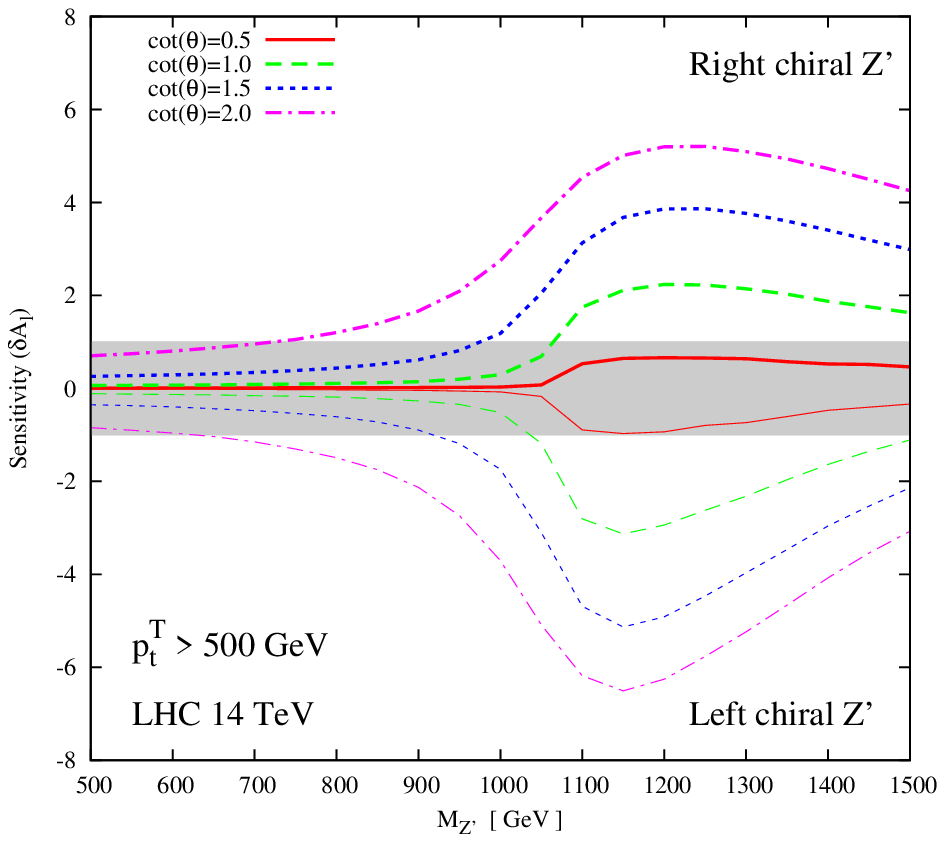, width=7.40cm}
\epsfig{file=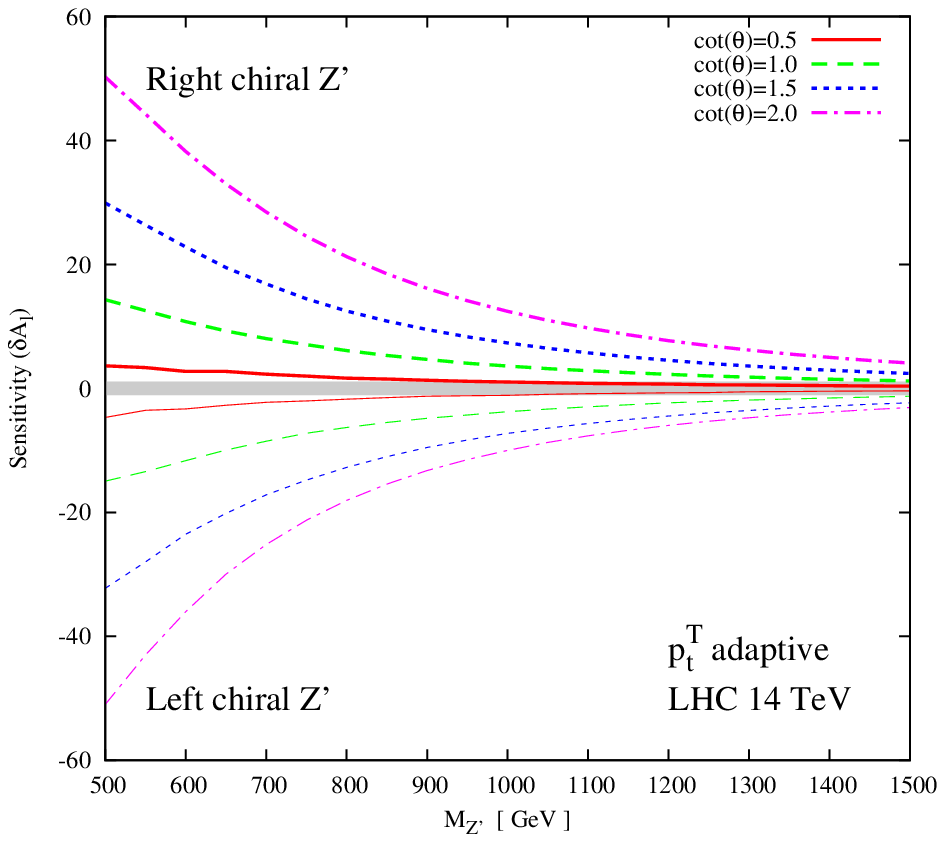, width=7.40cm}
\caption{\label{fig:sen_full_pt}The sensitivity for $\delta A_{\ell}$ as a function
of $M_{Z'}$ for $\sqrt{s} = 14$ TeV and integrated luminosity of $10$ 
fb$^{-1}$ ($\ell =e,\mu$). 
The shaded region corresponds to sensitivity between $-1$ and $+1$.
The legend is same as in Fig.~\ref{fig:al_full_pt}. } }
Similarly, in Fig.~\ref{fig:sen_full_pt}, we show the sensitivity for 
various $p_t^T$ cuts corresponding to the asymmetries shown in the 
Fig.~\ref{fig:al_full_pt}. 
For the right chiral $Z'$, the sensitivities are very large without any 
kinematical cuts.  This means that the use of
full set of events, without any cut, would be the best way to probe any new 
physics whose dynamics is expected to yield positive top polarizations
\cite{Djouadi:2007eg}. For 
negative top polarizations, however, one needs to use  kinematic cuts. For the
present case of $Z'$ the signal for negative top polarization can be 
enhanced with a cut on $m_{t\bar t}$ to select the resonance. This can be
seen from 
Fig.~\ref{fig:sen_full_mtt50} on comparing the sensitivities for left chiral
$Z'$ in the two panels. If, however, we want the asymmetry to tell us the sign
of the top polarization correctly, we need to employ transverse momentum
cuts. The corresponding sensitivities are also sizable for a large range of 
$M_{Z'}$ and $\cot(\theta)$ values. The best sensitivity, however, is 
achieved by
means of an adaptive $p_t^T$ cut. This cut requires a prior knowledge of
the mass and some idea about the width of the resonance. Having this 
information at ones disposal one can use this cut to estimate the
top polarization accurately once one has calibrated 
$\delta A_{\ell}$ against top polarization.

\subsection{The role of kinematic cuts}
We study here in some detail the effect of the kinematic cuts and how these
lead to the monotonic behaviour of the azimuthal asymmetry with polarization, 
a property  necessary for it to be a faithful probe of polarization.

To see the effect of the $m_{t\bar t}$ cut vis-\' a-vis the  $p_t^T$ cut 
we look at the $\phi_{\ell}$ distribution
for different $m_{t\bar t}$ and $p_t^T$ slices for the SM and for 
$Z'$ of a given mass
with left or right chiral couplings. These distributions, normalized to 
the rate in
that slice, are shown in Fig.~\ref{fig:phil_pt_mtt}. 
Before we analyze these distributions it is 
instructive to note the relation between $m_{t\bar t}$ and $p_t^T$. 

For a fixed $m_{t\bar t}$ the transverse momentum varies from $0$ to a maximum
value given by $(p_t^T)^{max} = \beta(m_{t\bar t}^2) m_{t\bar t}/2$.
Thus each slice in $m_{t\bar t}$ is an average
over this range of transverse momenta. Conversely, a given 
$p_t^T$ slice corresponds to a range of invariant masses with a lower
limit of $m_{t\bar t}^{min} =2\sqrt{(p_t^T)^2 + m_t^2}$, but no upper limit. 
\FIGURE[!ht]{
\epsfig{file=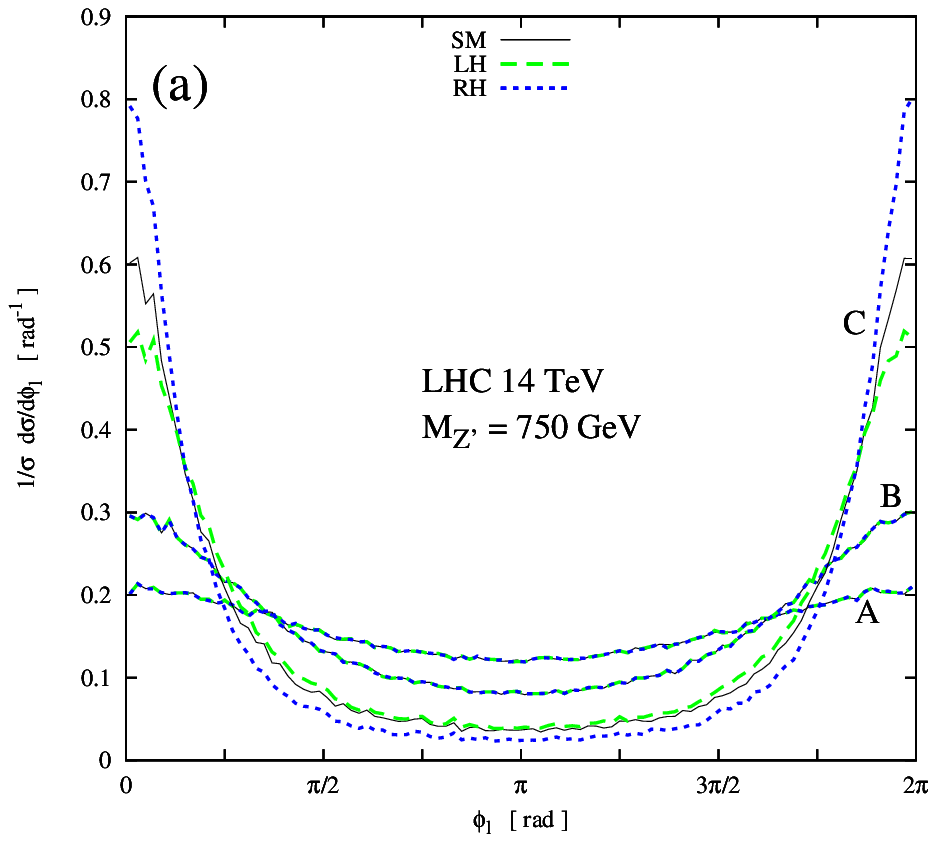, width=7.40cm}
\epsfig{file=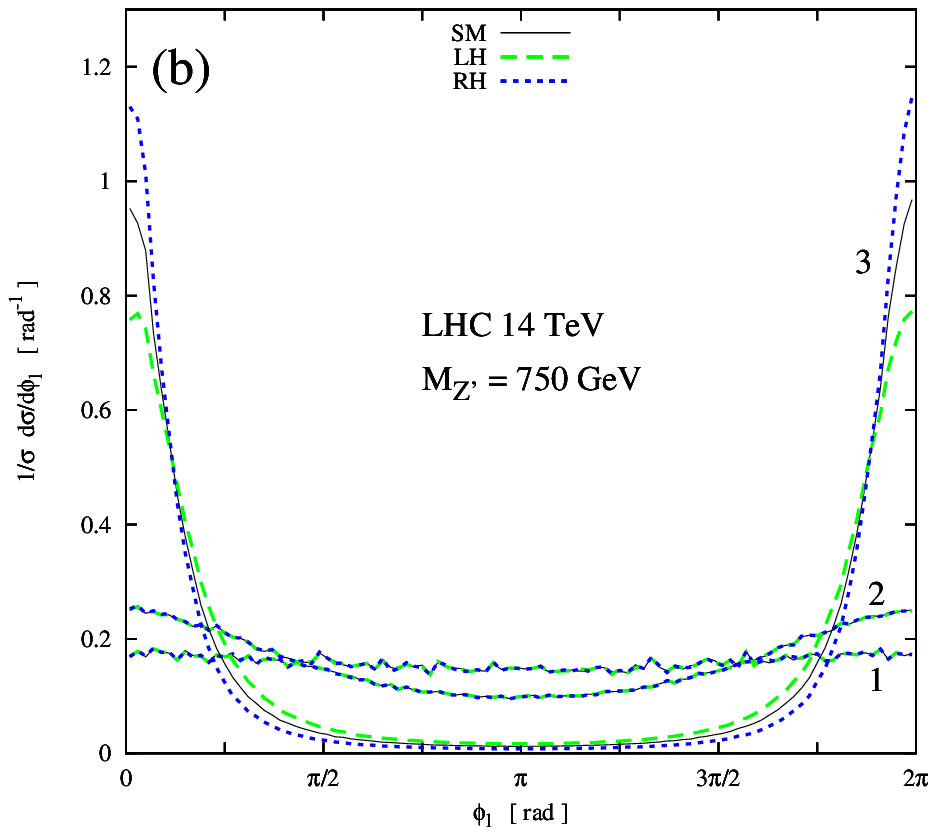, width=7.40cm}
\caption{\label{fig:phil_pt_mtt}
The normalized $\phi_{\ell}$ distributions for 
for the SM and $Z'$ 
of mass 750 GeV with both left and right chiral couplings (a) for
three different $m_{t\bar t}$ slices,
A$\equiv m_{t\bar t}\in [350,365]$ GeV,
B$\equiv m_{t\bar t}\in [395,410]$ GeV,
C$\equiv m_{t\bar t}\in [740,755]$ GeV, and  
(b)
for three $p_t^T$ slices,
$1\equiv p_t^T \in [0,15]$ GeV, $2\equiv p_t^T \in [45,60]$ 
GeV and $3\equiv p_t^T \in [330,345]$ GeV. }}

In Fig.~\ref{fig:phil_pt_mtt} (a), we show the $\phi_{\ell}$ distributions
for three different slices of $m_{t\bar t}$. The first one, A$\equiv 
m_{t\bar t}\in [350,365]$ GeV, is the lowest slice near the threshold of the
top pair production. Even for this slice the $\phi_{\ell}$ distribution is not flat.
This is due to the fact this slice contains a range of transverse
momenta
which can change the distribution. The second slice, B, is away from the 
threshold and also away from the $Z'$ resonance. Thus the slice B does not show
any sign of polarization, i.e. curves for the SM and for both chirality of the
$Z'$ the normalized distribution are identical. In the slice C, which is near
the $Z'$ pole, there is large change in the $\phi_{\ell}$ distribution owing to the
large top polarization near the resonance. But since the slice C also contains
events with a range of transverse momenta, this nice feature of the 
distribution may change if we look for heavier resonances. Already in the slice
C the curves for left and right chiral $Z'$ are not equidistant from the SM
curves at $\phi_\ell =0$, i.e. the transverse momentum dependent 
effect for the left chiral couplings is significant.

Next we look at Fig.~\ref{fig:phil_pt_mtt} (b), which shows the 
$\phi_{\ell}$ distributions for the same mass and couplings of $Z'$, but with 
different $p_t^T$ slices. The slice ``1'' is the lowest $p_t^T$ slice and the
$\phi_{\ell}$ distribution is flat owing to near zero polarization of the top sample
and the near zero transverse momentum. The second slice ``2'' has slightly 
higher transverse momentum and it starts to show the collimation near 
$\phi_{\ell}=0$ due to the effect of the transverse momentum. 
Since the top polarization for the slice ``2'' is also 
negligibly small, we have identical distributions for the SM and $Z'$ with both
chiralities. The slice ``3'' is near the peak in the $p_t^T$ distribution 
corresponding to the $Z'$ mass. Here we see a large collimation for all three
cases due to large $p_t^T$, yet the effect of the top polarization is clearly
visible. Also, the curves for left and right chiral $Z'$ appear to be 
equidistant from the SM curve for $\phi_\ell =0$.
The lepton asymmetry for this case will be a monotonically increasing
function of top polarization, as desired.

%

\subsection{Results for
lower energy colliders:
the Tevatron and LHC with $\sqrt{s}=7$ TeV}
\FIGURE[!ht]{
\epsfig{file=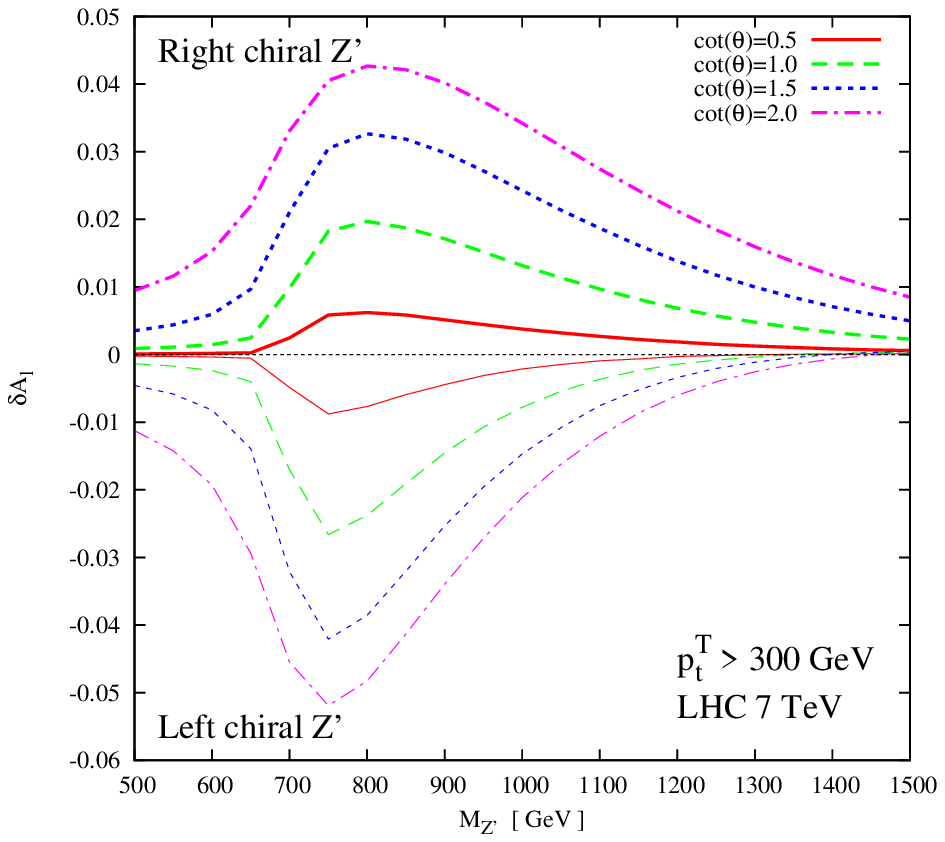, width=7.40cm}
\epsfig{file=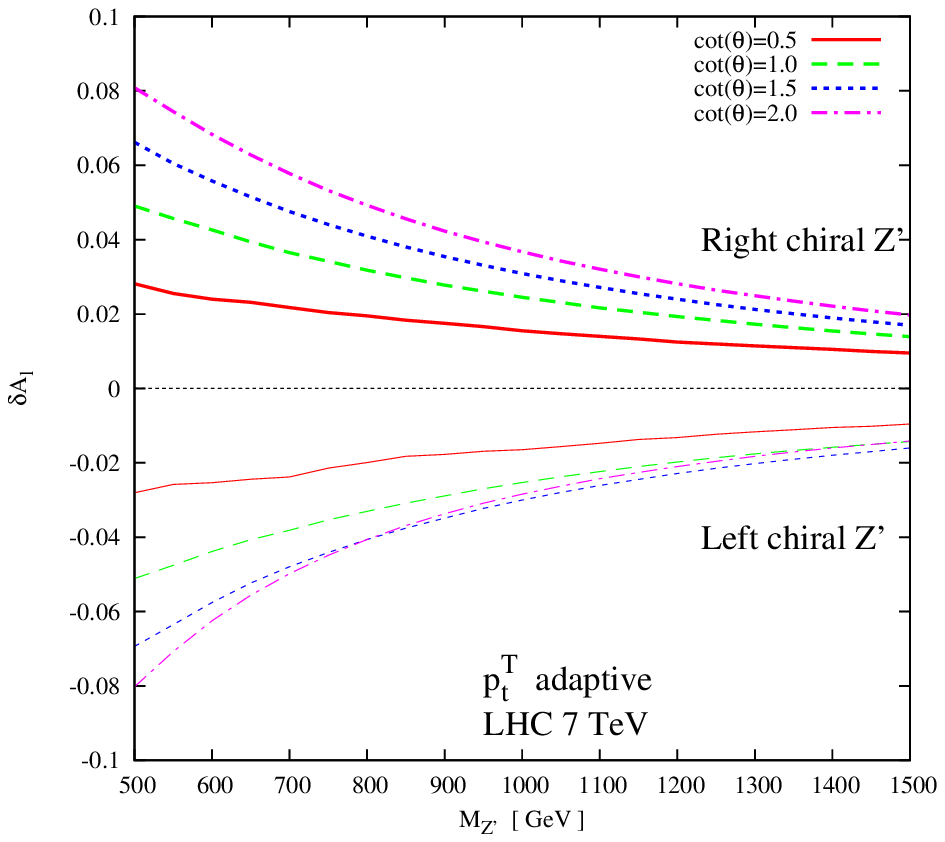, width=7.40cm}
\epsfig{file=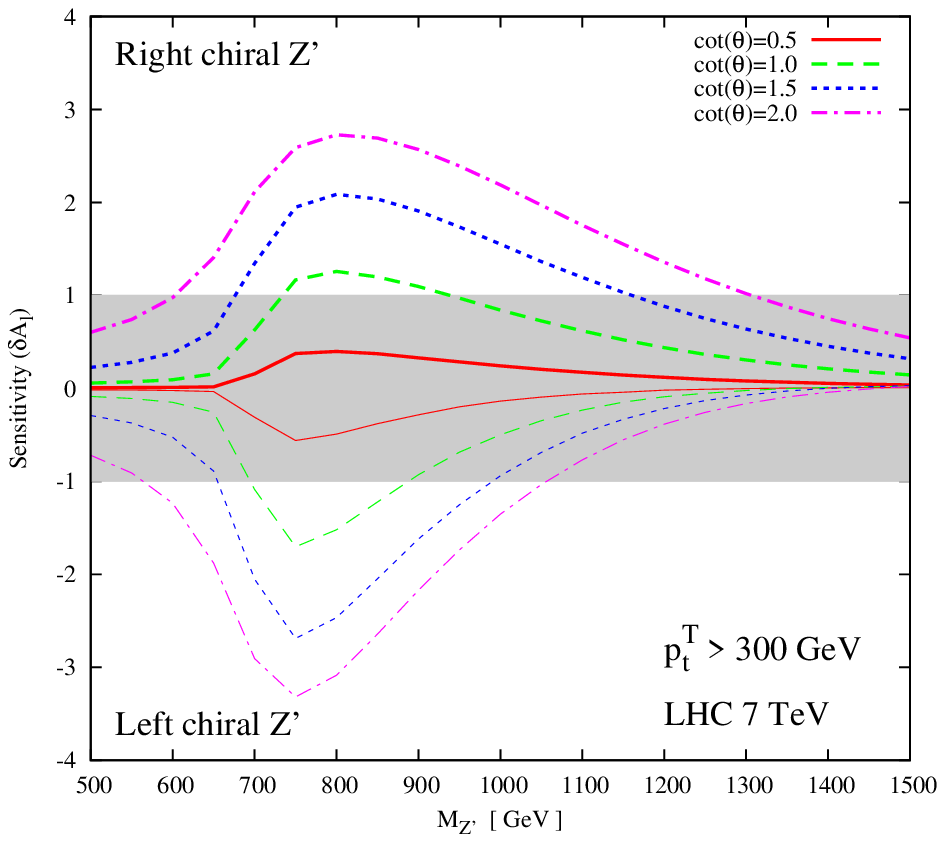, width=7.40cm}
\epsfig{file=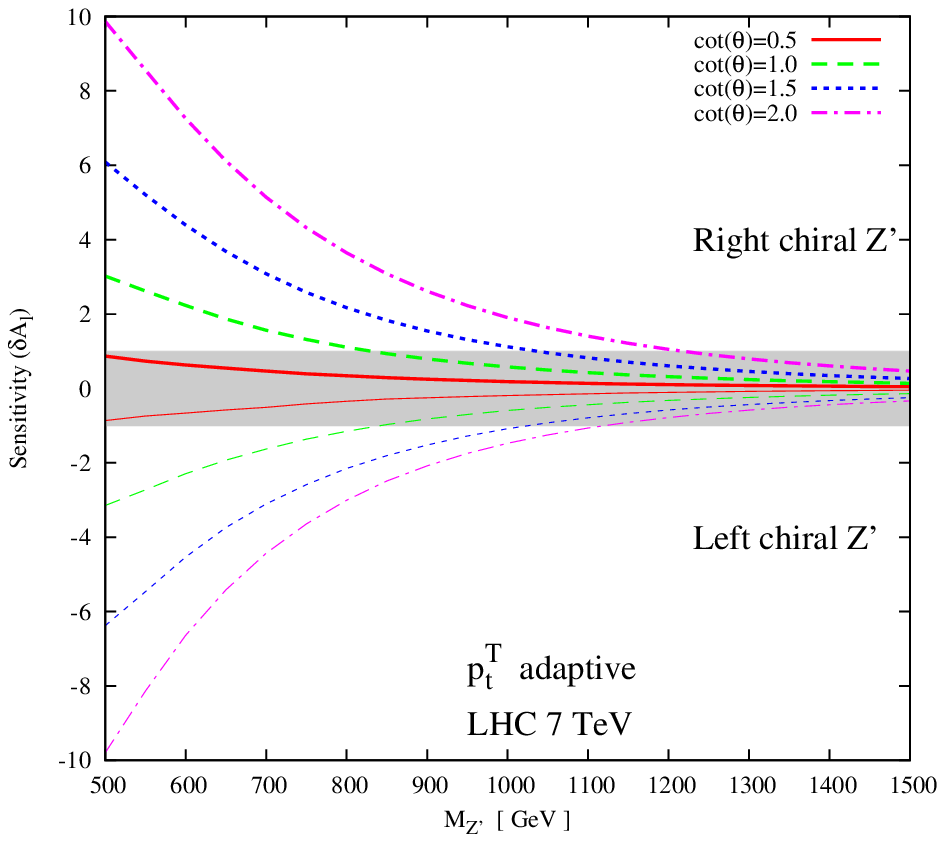, width=7.40cm}
\caption{\label{fig:LHC7}The asymmetry $\delta A_{\ell}$ as a function
of $M_{Z'}$ at the LHC with $\sqrt{s} = 7$ TeV with different $p_t^T$ cuts 
(top row)
and the corresponding sensitivity (bottom row) 
for integrated luminosity $1$ fb$^{-1}$ ($l=e,\mu$). 
The shaded region corresponds to sensitivity between $-1$ and $+1$.
The legend is the same as in 
Fig.~\ref{fig:al_full_mtt50}. } }
At present, both LHC and Tevatron are running. The LHC running at
$\sqrt{s} = 7$ TeV,
is expected to accumulate about $1$ fb$^{-1}$ integrated luminosity before 
upgrading to higher energy. The Tevatron is expected to accumulate a total
of $15$ fb$^{-1}$ before shutting down. Thus it will be
instructive to look at the lepton asymmetry and corresponding sensitivities 
for these two collider options.
For LHC at $\sqrt{s} = 7$ TeV, we show the asymmetry $\delta A_{\ell}$ for two 
different cuts,
fixed cut of $p_t^T>300$ GeV and adaptive $p_t^T$ cut in the top row of 
Fig.~\ref{fig:LHC7}.
The corresponding sensitivities are shown in the bottom row of the same figure.
Similarly, Fig.~\ref{fig:TEV} shows the asymmetry and the corresponding 
sensitivities for the Tevatron.

Due the lower energy and low luminosity at the $\sqrt{s} = 7$ TeV and
$\int L dt = 1$ fb$^{-1}$ run of the LHC, 
the sensitivity is $7-8$ times smaller than that for the LHC at
$\sqrt{s}=14$ TeV and $\int Ldt =10$ fb$^{-1}$.
On the other hand, it is comparable to that for the Tevatron with $\int L
dt = 15$ fb$^{-1}$ with the
fixed $p_t^T$ cut and smaller than that for the Tevatron run for the 
adaptive $p_t^T$ 
cut. The reason is that Tevatron being  a $p\bar p$ machine, the $q\bar q$
luminosity is higher than at the LHC. 
Further, the lower energy of the Tevatron leads to a reduced background from 
$gg\to t\bar t$ as compared to the LHC 7 TeV run. Hence the 
Tevatron is more sensitive
than LHC at $\sqrt{s} =7$ TeV with the adaptive $p_t^T$ cuts.
It should be remembered that at the Tevatron the existence of a unique
definition of the $z$ axis, might offer us the possibility of constructing
additional observables/asymmetries using the polar angle of the $\ell$ as well.
This will be discussed elsewhere.
\FIGURE[!ht]{
\epsfig{file=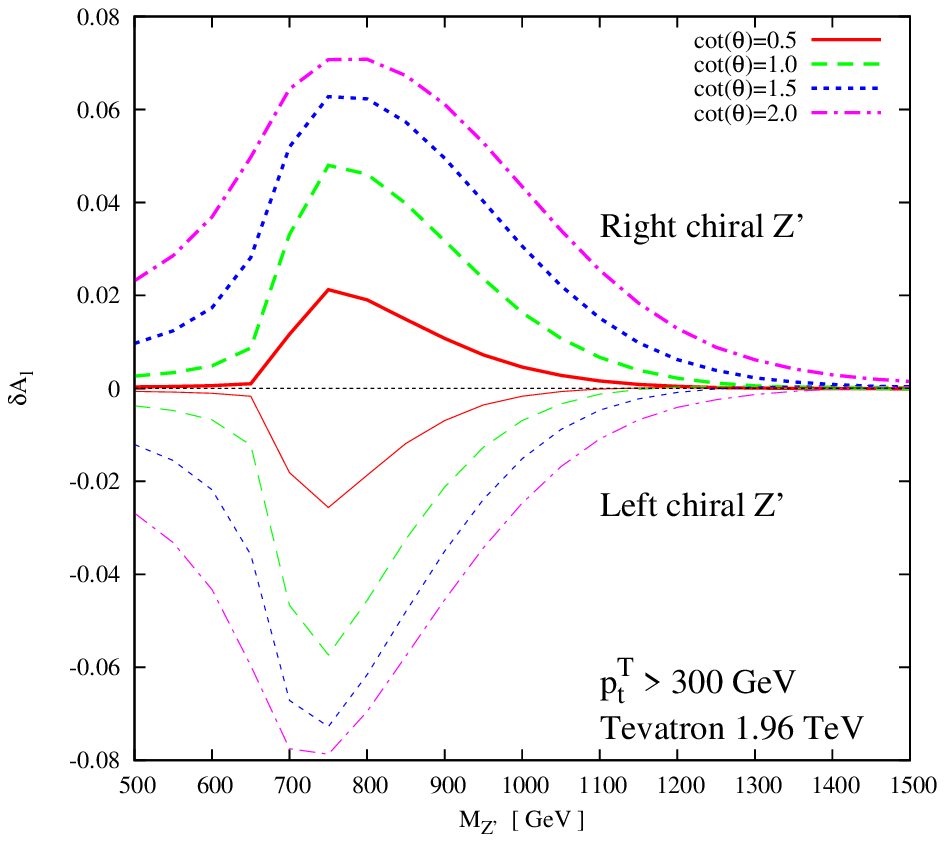, width=7.40cm}
\epsfig{file=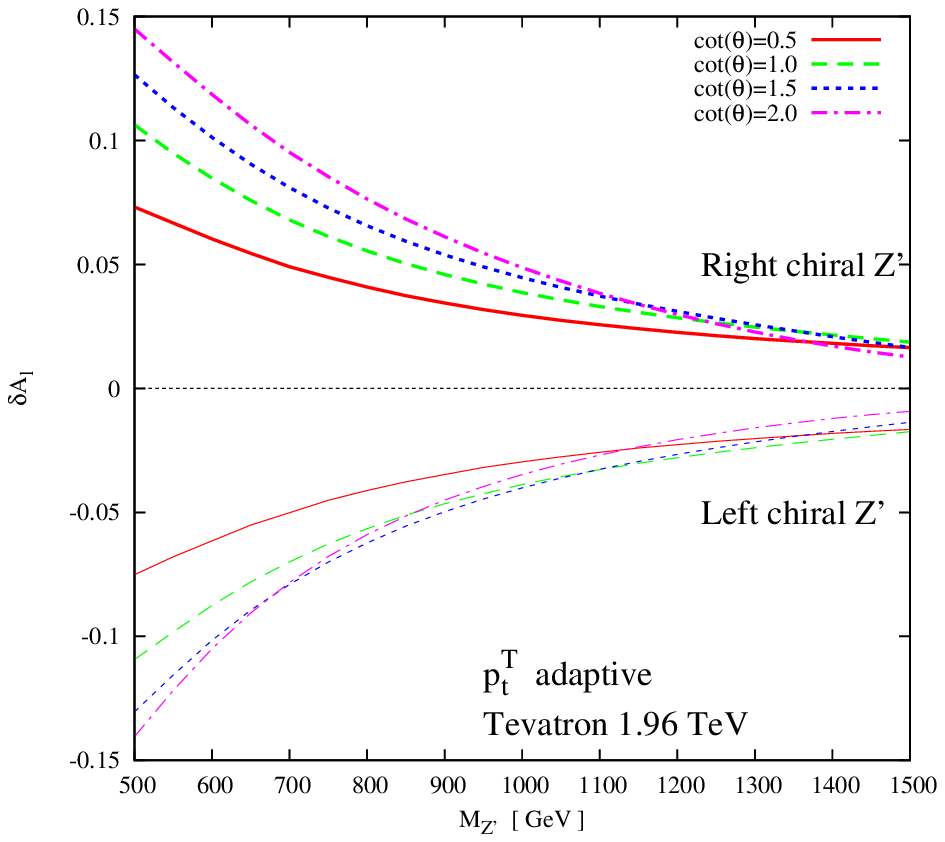, width=7.40cm}
\epsfig{file=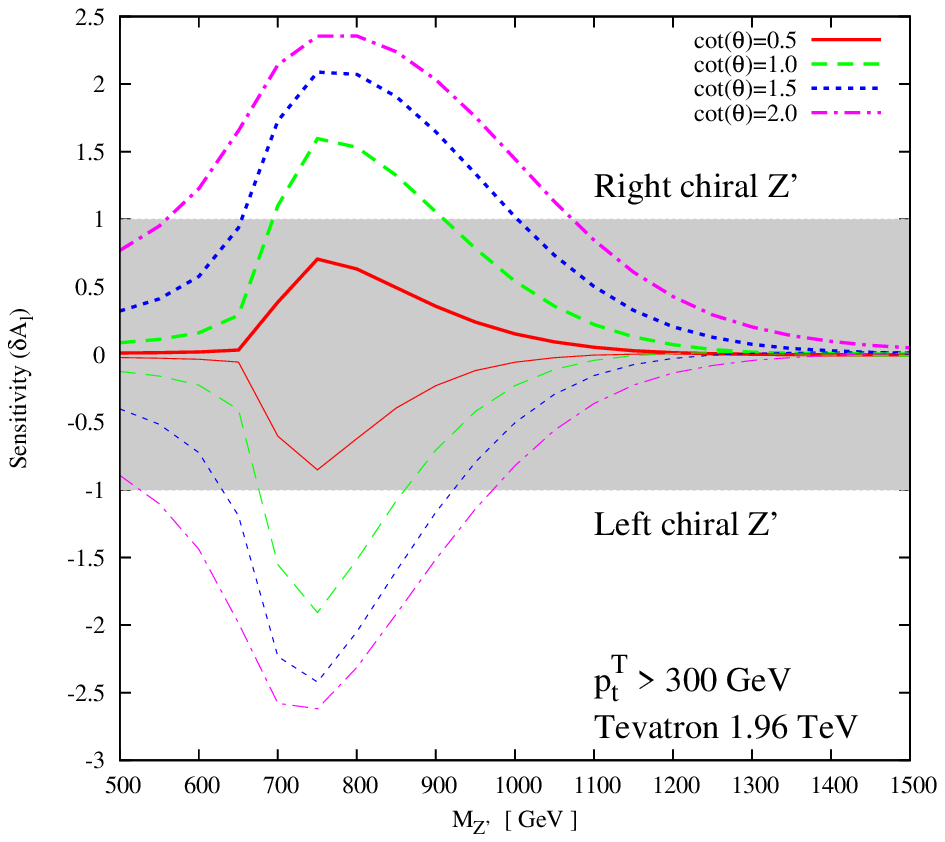, width=7.40cm}
\epsfig{file=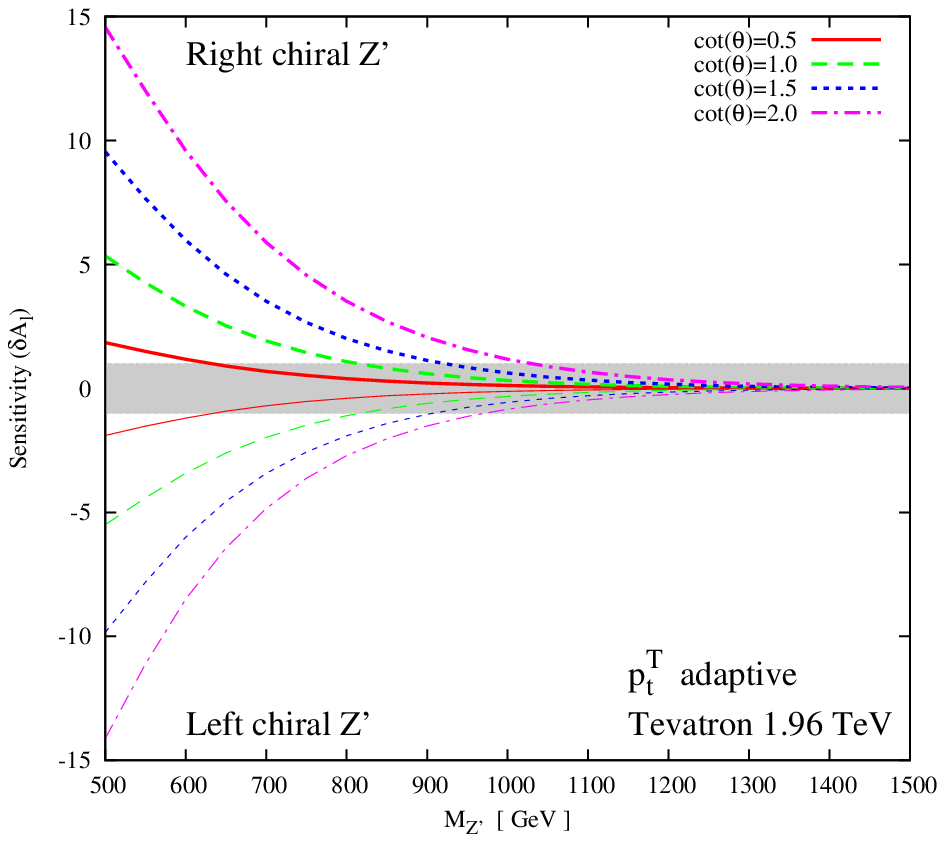, width=7.40cm}
\caption{\label{fig:TEV}The asymmetry $\delta A_{\ell}$ as a function
of $M_{Z'}$ at the Tevatron 
with $\sqrt{s} = 1.96$ TeV 
with different $p_t^T$ cuts 
(top row)
and the corresponding sensitivity (bottom row) for 
integrated luminosity of $15$ fb$^{-1}$ ($l=e,\mu$). 
The shaded region corresponds to sensitivity between $-1$ and $+1$.
The legend is the same as
in Fig.~\ref{fig:al_full_mtt50}. } }

\section{Conclusions}
In this note we have investigated use of {\it single top polarization}
as a probe of the $t \bar t$ production mechanism. To that end we have 
constructed an observable, which can reflect the sign and the magnitude 
of the top polarization faithfully. We do this by using  the azimuthal 
angle distribution of the decay lepton in the laboratory frame which carries
information on the top-quark polarization. For purposes of illustration we have
chosen a concrete model, inspired by the Little Higgs models, 
which has an additional spin-1 boson $Z'$ with mass $M_{Z'}$ and  
chiral couplings to the quarks.  In addition to the mass $M_{Z'}$,
this  model is  characterized by one more parameter, $\cot(\theta)$
which gives the strength of the couplings.
\comment{
We first studied the 
cross sections for producing a $t\bar t$ pair, where only the top has a
definite helicity, as also the degree of polarization of this top, as a
function of the $t\bar t$ invariant mass $m_{t\bar t}$.}
We began by studying the cross sections for producing a $t \bar t$ pair,
where the top quark has a definite helicity (the cross-section being summed
over the helicity state of the anti-top) and hence the degree of polarization
of the produced top, as a function  of the $t\bar t$ invariant mass 
$m_{t\bar t}$.  We
find that the top polarization dependent part of the cross section in the
model can be large, even comparable to the unpolarized cross section, 
 in the region of the $Z'$ resonance. We also calculated the degree of
top polarization in the model, and found it to be of the order of a few per
cent for $Z'$ masses around 1000 GeV, and larger for lower $M_{Z'}$ (for 
example it has a value $\sim 10\%$ for a $Z'$ with mass $700$ GeV for $\sqrt{s} 
= 7$ TeV) as compared to a value of less than $10^{-3}$ expected in the SM. 
The sign of the polarization follows the chirality of the $Z'$
couplings to $t\bar t$. Further the polarization dependent part of the cross
section was also found to peak in the region of the top-quark
transverse momentum $p_t^T \approx \sqrt{1-4 m_t^2/M_{Z'}^2}M_{Z'}/2$.
Hence the $t$ polarization can be maximized by appropriate cuts on
$m_{t\bar t}$ or $p_t^T$.

We then investigated to what extent the azimuthal angular distribution of
the charged lepton produced in top decay would mirror the extent of this
large top polarization. It turned out that without any cuts, the
normalized azimuthal distribution is sensitive to the magnitude and sign 
of the top polarization only for small $M_{Z'}$, up to about 600 GeV. 
The top polarization modifies the height of the peak that this distribution
has  near 
near $\phi_{\ell} \approx 0$ (and $\phi_{\ell} \approx 2\pi$ ).
The peak is higher (lower)  for right 
(left) chiral couplings than for the SM: for example for $M_Z = 500$ GeV a
polarization of about $12\%$ caused the peak to be higher by about $10 \%$.
This distribution is not symmetric in $\cos\phi_\ell$, and we can define
an asymmetry  $A_\ell$ about $\cos\phi_\ell =0$.
Since the initial state at the LHC has identical particles, choosing the
beam axis as the $z$ axis does not allow for a unique choice of the
direction in which $z$ is positive, leading to distributions which are
symmetric under $\phi_\ell \rightarrow 2 \pi - \phi_\ell$. This does not
preclude, however, an asymmetry of the azimuthal distribution about
$\cos\phi_\ell = 0$.
$A_\ell$ has a nonzero value, $A_\ell^{SM}$, for the SM, i.e. the case
of an unpolarized top. The deviation of $A_\ell$ from its SM value,
$\delta A_\ell$,  is sensitive to $\cot(\theta)$, as well as to the chirality 
of the $Z'$ couplings for $M_{Z'} < 600$ GeV.  We observe, however, that 
$\delta Al$ becomes positive, for larger values of $M_{Z'}$,
irrespective of the chirality. This indicates that the
azimuthal distributions and asymmetry get contributions which are partly
dependent on the top polarization, and partly purely kinematic in nature.

We then investigated effects of kinematic cuts in order to make the 
$\delta A_{\ell}$ more faithful to the sign and the magnitude of the $t$ 
polarization, and hence to the couplings of the $Z'$, for a larger range 
of $M_{Z'}$.  A cut on $m_{t\bar t}$ restricting it to the resonant
region around $M_{Z'}$ makes the $\delta A_\ell$ independent of  $M_{Z'}$,
for the full range considered for right chiral couplings. Even though for
the left chiral couplings this still happens only for a limited
range of $M_{Z'}$, the  range is now larger than without any cuts. Thus
knowing the mass of the resonance will already be of help.
A cut on the top transverse momentum $p_t^T$ restricting it to values
larger than a fixed value of a few hundred GeV succeeds in getting 
$\delta A_\ell$ to reflect faithfully both the magnitude and 
the chirality of the coupling, alternatively magnitude and 
sign of the $t$ polarization, irrespective of $M_{Z'}$.
An adaptive cut, in which $p_t^T$ is restricted to a window which depends
on the width of $Z'$, and hence on the coupling $\cot(\theta)$, makes
$\delta A_\ell$ more sensitive to lower values of $M_{Z'}$, even though not
completely monotonic in $M_{Z'}$. Interestingly, now for the polarization 
values of a few per cent one gets $\delta A_\ell$ of the same order, and even
enhanced by a factor of $2-3$ by the adaptive $p_t^T$ cut. 

The statistical significance of the azimuthal asymmetry for various
kinematic cuts was also examined, with the conclusion that the
sensitivity is large for all values of $\cot(\theta) > 0.5$ that we
consider. The best sensitivity is achieved with the use of the
adaptive $p_t^T$ cuts. As an example, for the design energy of the
LHC of $\sqrt{s}=14$ TeV and an integrated luminosity of 10 fb$^{-1}$,
even with a plain cut on $p_t^T$ we find sensitivity values $\geq 3$
over a large part of the range of $M_{Z'}$ values considered, extending to
large $M_{Z'}$. 

We also evaluated the sensitivity of our observable for the current run 
of LHC with $\sqrt{s}= 7$ TeV and integrated luminosity of 1 fb$^{-1}$, 
as well as for the Tevatron with an integrated luminosity of 15 fb$^{-1}$. 
At $\sqrt{s} = 7$ TeV values of asymmetry $\delta A_{\ell}$ as high as
$4$--$5 \% $ ($6$--$7 \%$) for a fixed (adaptive) $p_t^T$ cut can be
reached. Due to the smaller luminosity for this run, the sensitivity values
are rather low and above $1\sigma$ only for $M_{Z'}$ values between
$800$ to $1200$ GeV and for larger values of $\cot(\theta) $. It was found 
that the sensitivity  at the Tevatron could be comparable to that at the 
LHC with $\sqrt{s} = 7$ TeV, though less than that at the 14 TeV version 
of the LHC.
 
In conclusion, the leptonic azimuthal asymmetry, with suitable cuts can be 
a useful tool for studying mechanisms for top production which can give rise 
to large spin-dependent effects. 
\acknowledgments
R.G. wishes to acknowledge support from the Department of Science and 
Technology, India under Grant No. SR/S2/JCB-64/2007, under the J.C. Bose
Fellowship scheme. K.R.  gratefully acknowledges support from the
Academy of Finland (Project No. 115032). S.D.R. thanks Helsinki Institute
of Physics for hospitality during the period of completion of this work.
The work of R.K.S has been supported by the German Ministry of Education and
Research (BMBF) under contract no. 05H09WWE.


\end{document}